\documentclass[sigconf]{acmart} 

\AtBeginDocument{%
  \providecommand\BibTeX{{%
    \normalfont B\kern-0.5em{\scshape i\kern-0.25em b}\kern-0.8em\TeX}}}

\copyrightyear{2025}
\acmYear{2025}
\setcopyright{acmlicensed}\acmConference[IMX '25]{ACM International Conference on Interactive Media Experiences}{June 3--6, 2025}{Niterói, Brazil}
\acmBooktitle{ACM International Conference on Interactive Media Experiences (IMX '25), June 3--6, 2025, Niterói, Brazil}
\acmDOI{10.1145/3706370.3727867}
\acmISBN{979-8-4007-1391-0/2025/06}

\usepackage{CJKutf8}
\usepackage{array}
\usepackage{tabularx}
\usepackage{float}
\usepackage{balance}

\usepackage{array}

\newcolumntype{M}[1]{>{\centering\arraybackslash}m{#1}} 

\newcolumntype{L}[1]{>{\raggedright\let\newline\\\arraybackslash\hspace{0pt}}m{#1}}
\newcolumntype{C}[1]{>{\centering\let\newline\\\arraybackslash\hspace{0pt}}m{#1}}
\newcolumntype{R}[1]{>{\raggedleft\let\newline\\\arraybackslash\hspace{0pt}}m{#1}}
\begin{document}

\title[VIBES]{VIBES: Exploring Viewer Spatial Interactions as Direct Input for Livestreamed Content}

\author{Michael Yin}
\affiliation{
  \institution{University of British Columbia}
  \city{Vancouver}
  \state{BC}
  \country{Canada}
  \postcode{V6T 1Z4}
  \orcid{0000-0003-1164-5229}
}
\email{jiyin@cs.ubc.ca}

\author{Robert Xiao}
\affiliation{
  \institution{University of British Columbia}
  \city{Vancouver}
  \state{BC}
  \country{Canada}
  \postcode{V6T 1Z4}
  \orcid{0000-0003-4306-8825}
}
\email{brx@cs.ubc.ca}

\begin{abstract}
Livestreaming has rapidly become a popular online pastime, with real-time interaction between streamer and viewer being a key motivating feature. However, viewers have traditionally had limited opportunity to directly influence the streamed content; even when such interactions are possible, it has been reliant on text-based chat. We investigate the potential of spatial interaction on the livestreamed video content as a form of direct, real-time input for livestreamed applications. We developed VIBES, a flexible digital system that registers viewers’ mouse interactions on the streamed video, i.e., clicks or movements, and transmits it directly into the streamed application. We used VIBES as a technology probe; first designing possible demonstrative interactions and using these interactions to explore streamers' perception of viewer influence and possible challenges and opportunities. We then deployed applications built using VIBES in two livestreams to explore its effects on audience engagement and investigate their relationships with the stream, the streamer, and fellow audience members. The use of spatial interactions enhances engagement and participation and opens up new avenues for both streamer-viewer and viewer-viewer participation. We contextualize our findings around a broader understanding of motivations and engagement in livestreaming, and we propose design guidelines and extensions for future research. 
\end{abstract}



\begin{teaserfigure}
\centering
  \includegraphics[width=0.85\textwidth]{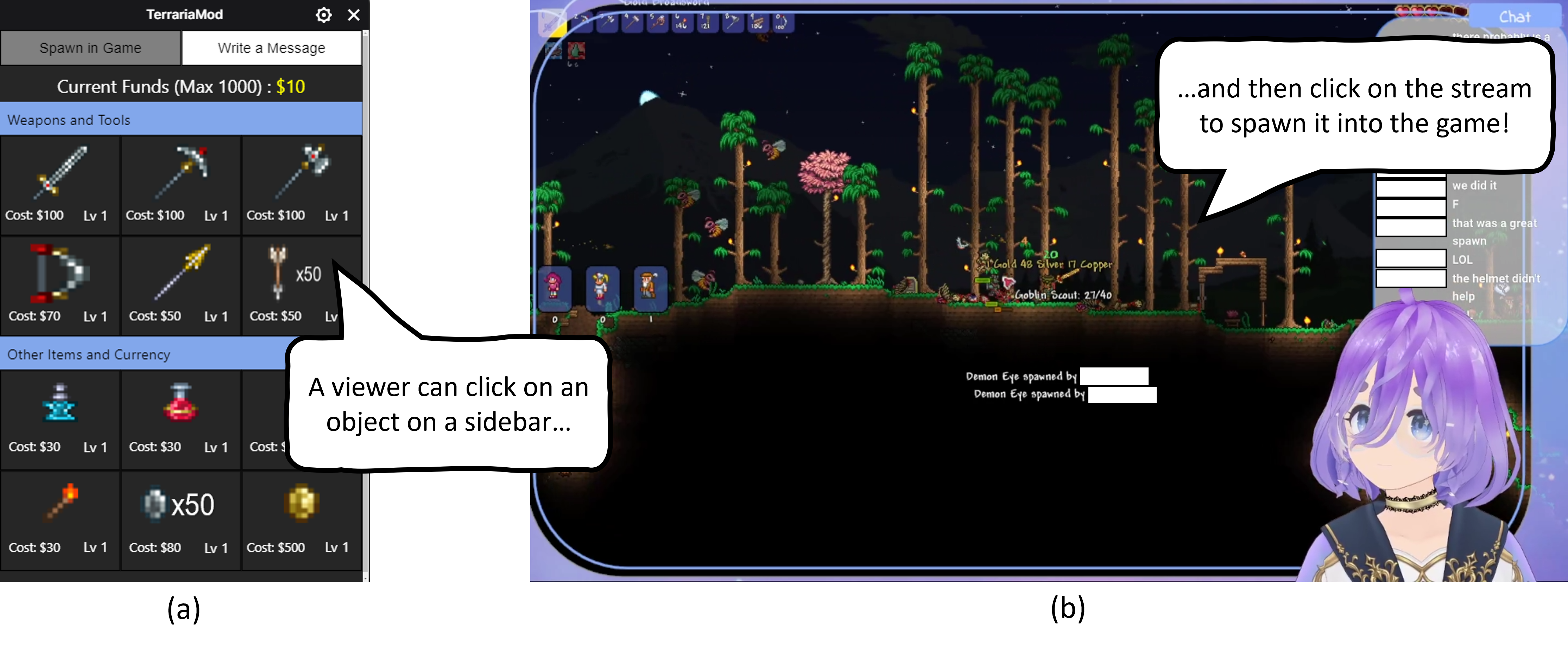} 
  \caption{VIBES allows audience members of a livestream to spatially interact directly with the streamed application through mouse events on the video player. In this example, viewers can spawn items into a game by first clicking on the item on the Twitch extension \emph{(a)} and then clicking on the desired spawn location on the streamed game \emph{(b)}.}
  \Description{The first subimage displays a Twitch extension, which is an iframe embedded within the Twitch video player. On the iframe are pictures of items that users can click to spawn into the streamer's game, such as tools, weapons, and currency. The second subimage depicts the actual stream. In the stream, enemies are being spawned in by viewers and the streamer is fighting against these spawned enemies.}
  \label{fig:teaser}
\end{teaserfigure}

\maketitle

\section{Introduction}
\textbf{Livestreaming} (or simply \textbf{streaming}) is a massive media phenomenon in which streamers broadcast their activities in real-time to audience members around the world \cite{Taylor+2019, SJOBLOM2017985}. Streaming platforms, such as Twitch and Youtube Live, provide the means of distribution for a streamer to share content \cite{wohn2020live, alexander2021}. The appeal of livestreaming has been described as transforming “private play into public entertainment”, in which the streamer, the audience, and their interactions all become part of the spectacle \cite{Taylor+2019}. This idiosyncratic element of mixing real-time interactivity with performative elements has had a profound influence on academic research \cite{SJOBLOM2017985, li2020, wohn2018, li2019coperformance}. Research into motivations behind livestreaming, from both viewer and streamer perspectives, highlights the importance of social interactions \cite{li2020, xu2020watches, gros2017, lottridge2017, SPEED2023100545} as a unique form of motivation for engagement \cite{Taylor+2019}.

However, viewer-streamer interaction methods remain rather limited in practice, with the primary facilitating medium being text-based chat \cite{li2020, smith2013, striner2021, robinson2022, chung2021}. To address these constraints, streaming platforms have introduced additional interaction mechanisms such as subscriptions, gifting, or donation systems \cite{chen2021gifting, Li2021}. Explorations have also looked into expanding the degree of influence of chat messages, for example, having direct effects on the streamed content itself as in \emph{Twitch Plays Pokemon} \cite{lessel2017, ramirez2014twitch} or \emph{Marbles on Stream} \cite{Marbles}. Recently, researchers have extended these ideas by investigating how non-textual forms of viewer interaction may affect the content of the stream, further broadening the scope of viewer participation and engagement \cite{chung2021, robinson2022}.

Building on this prior research, our exploratory work investigates the broad input method of \textbf{spatial interaction events}. To probe upon this concept, we scope our research specifically on mouse events (clicks and motion) on the streamed video as a direct input mechanism, leveraging existing elements within browser-based livestreams. Unlike text-based chat, mouse events can encode \textbf{direct spatial input} that maps onto the visual content of the livestream, enabling users to interact \emph{through} the stream rather than \emph{about} it. Our research aims to explore two research questions:

\begin{itemize}
    \item \textbf{RQ1} --- How does direct spatial input affect the streamer’s experience, and what challenges and opportunities emerge for streamers?
    \item \textbf{RQ2} --- How do viewers perceive and engage with direct spatial input, and how does it influence their interactions within the stream?
\end{itemize}

To address these, we developed \textbf{VIBES} (\textbf{V}ideo-based \textbf{I}nteractions for \textbf{B}roadcasted \textbf{E}vents in \textbf{S}treaming), a lightweight, flexible prototype that collects and deciphers users’ mouse-based interaction events and transmits them to livestreamed applications. VIBES was a technology probe \cite{techprobe} --- a simple, adaptable system that allowed us to gain insights into user and streamer experiences using such events. An initial formative exploration illuminated streamer perspectives on how audience spatial input could integrate with their motivations to stream and affect their streaming experiences. A subsequent user study involving the deployment of VIBES in livestreams highlighted practical insights and the effect of direct spatial input on viewer and streamer engagement (addressing \textbf{RQ2}). Our findings extend prior research on livestreaming engagement and motivations, and we highlight challenges and present suggestions for future exploration. 

\section{Related Works}
We considered prior research to motivate the research gap, inform the development of our probe, and contextualize potential motivational and engagement impacts on streamers and viewers. 

\subsection{Motivations and Relationships --- Viewer and Streamer Perspectives}
We considered the foundational motivations for streaming and viewership. From a streamer perspective, Taylor argued that motivations for streaming stem from social connections --- the transformation of play experiences into public performances \cite{Taylor+2019}; Li et al. found that streamer motivations comprise reasons relating to social integration, personal integration, and affection \cite{li2020}. Similar motivations have been found in studies regarding educational \cite{chen2021} and cultural streaming \cite{lu2019} --- social engagement through connecting with viewers forms a major part of why streamers stream. 

For viewers, streams satisfy desires for social contact and entertainment. The real-time aspect allows for personal, intimate connections with the streamer \cite{lin2017, wohn2020} and other viewers \cite{lu2019, hamilston2016} --- being alongside others in a shared community forms this social motivator \cite{Taylor+2019, lu2018}. Li et al. \cite{li2020} found that viewer identification and streamer relationship were core facets underlying the motivation for viewership; Hamilton et al. found that direct recognition by the streamer can be a rewarding experience for viewers \cite{hamilton2014}. Sjöblom and Hamari considered viewer motivations from a uses and gratifications perspective, finding that viewership was correlated to contextually translated human needs, such as acquiring information, enhancing connections, and escape \cite{SJOBLOM2017985}. From a viewer's perspective, social engagement with the streamer and the other viewers extends the underlying entertainment factor as a major reason to watch streams. 

Common in both streamer and viewer motivations is the high degree of focus on bi-directional sociality --- as livestreaming is a real-time activity, it promotes a high degree of synchronous engagement \cite{li2020, zimmer2019drives}. In the act of streaming, the audience and their interactions become integrated into the entire experience and become “part of the show” \cite{Taylor+2019}. Smith et al. stated that the interaction between these two agents can be “entertaining and reciprocal” \cite{smith2013}, and that viewers can help co-create content. Hamilton et al. expanded on how viewers want to interact with the stream, and streamers often give opportunities for viewers to do so beyond simply chatting, such as having them play games with the streamers \cite{hamilton2014}. Our work evaluates how providing users with a way to directly impact the streamed application using their visuospatial mouse interaction events affects this co-creation of entertainment through the lens of streamer and viewer engagement. We consider both how this affordance affects the present level of communication and engagement in the streamer-viewer relationships and how it might encourage new forms of social interaction. 

\subsection{Performative Audience Participation and Collaboration} 

Past research has investigated methods of audience participation within streaming. Interactive interactions with video-based media have long existed, such as collaborative annotation of online videos \cite{cross2014vidwiki, leetiernan2001fostering}. Outside the digital realm, Webb et al. coined the term “distributed liveness”, which comprises the real-time relationships between performers and audiences \cite{webb2016}. The idea of having audience members directly influence real-time performance and artistry has been explored in many fields, such as theatre, \cite{Cerratto-Pargman}, orchestra \cite{roberts2011composition}, VR experiences \cite{Herscher2019CAVRN}, and video meetings \cite{tee2006}. 

Livestreaming opens up new opportunities regarding distributed liveness, especially due to the asymmetric nature of audience participation at scale. This has been considered in audience-participation games (APGs), which allow audience members to impact gameplay in significant ways \cite{Seering2017}. One example of an APG is Twitch Plays Pokemon (TPP), in which audience members were given decentralized control over the button input of a livestreamed game \cite{lessel2017}; this later spawned the formation of a community and surrounding narrative \cite{ramirez2014twitch}. The impact of TPP has been explored in academic literature, and findings suggest that it creates unique forms of social agency \cite{Seering2017}. More generally, Striner et al. presented a theme map of audience participation in livestreaming based on student designers, where concepts such as streamer-viewer relationships, interaction, and agency were key \cite{striner2021}.

APGs allow audience members to encapsulate individual and collective roles within the game itself, opening up new methods of interactions \cite{Seering2017}. User integration into the streamed game affords engagement, but the medium of chat limits the ability for viewer expressiveness to flourish \cite{lessel2017}. This indicates a possible research gap --- what are ways to directly engage with the content in more expressive manners than text-based chat? We hark towards research comparing GUIs and CLIs. For GUIs, input to a device is passed through interaction events such as click or touch events, which encode spatial expression; certain implementations of such systems have had advantages over text-based CLIs in terms of user satisfiability and effectiveness \cite{staggers_comparing_2000, a_feizi_usability_2012}. An analogous implementation for interactive livestreaming is to use similar events as input into the streamed content, which we investigate in this work. This can relate to theories drawn from psychophysiological foundations, understanding how different sensorimotor signals (such as vision and movement) can have different responses and outcomes, such as affective state \cite{goriMultisensoryPerceptionLearning2022, ravajaContributionsPsychophysiologyMedia2004}. 

A related variant to APGs are danmaku-participation games (DPG) \cite{wangLetsPlayTogether2023a}, in which viewers can use danmaku as a form of control in a streamed game; these games have often been studied without the presence of an actual streamer \cite{wangLetsPlayTogether2023a, chenFeltEveryoneWas2024}. Danmaku is a popular audience interaction paradigm on Asian video sites in which people's textual comments float across the screen, often used for expression, entertainment, and social communication \cite{zhaoExploringMotivationalAffordances2016, liuWhoYouIntegrating2017, panMotivationsGameStream2023}. DPGs, like APGs, provide audience members with control over the streamed application through their danmaku input, as well as offer similar motivations, such as engagement and connection, with similar challenges, such as latency and instability \cite{wangLetsPlayTogether2023a}.

Persistent obstacles for APGs include latency, screen real-estate, and player agency \cite{glickman2018}. These are all aspects we consider in our probe. Regarding agency --- unlike other systems such as StreamSketch \cite{lu2021streamsketch}, which allow user input on the video to be visualized for the streamer but puts the onus of decision-making with the streamer; we study the fundamental transfer of agency towards the audience, as their inputs can influence the application as much as the streamer's inputs. Regarding latency, we directly address this issue of latency in development, and regarding screen real-estate, we re-purpose components that already exist in livestreams. Ultimately, in our work, we consider how visual input through direct spatial input capturing interactive events modulates existing learnings on audience participation and interaction within livestreaming.

\subsection{Beyond Just Chatting --- Livestreaming Input Methods}

Researchers and developers have built systems that extend the prevalent paradigm of chat interaction on livestreams and video content. Some of these systems are integrated directly into the platform, such as polling or gifting systems \cite{chen2021gifting, Li2021}. Others are more exploratory, such as using audience physiological markers \cite{robinson2022, huangSharingFrissonsOnline2024a}, voice messages \cite{ahn2022}, or haptic feedback \cite{kim2018} to affect content. For instance, Huang et al. studied the use of viewer frissons to affect video content in increasing the social affective dimension \cite{huangSharingFrissonsOnline2024a}; Robinson et al. studied how using audience heartbeat to influence gameplay resulted in a positive boon to engagement \cite{robinson2022} --- our study follows a similar evaluation as the latter to assess the impact of VIBES. Furthermore, Fanzo et al. built a game that asymmetrically allows users to view different camera channels within a game \cite{fanzo2017}, and Lessel et al. performed case studies on an audience communication extension that allows polling and tracking in Hearthstone \cite{lessel2017expanding} and aggregation of move-selection for chess \cite{crowdchess}. 

Hammad et al. developed MARS, which augments the stream's video player with a context-aware visualization of game metadata to support novel game designs and provide added information to the viewers \cite{Hammad2023MARS}; however, it does not necessarily support input from the viewers in affecting the stream itself. Yang et al. developed Snapstream, which allows users to create snapshots of the stream and share these annotated snapshots \cite{yang2020snapstream}. The use of interactive visual elements based on the streamed content was found to increase engagement and communicative expression. The importance of visual modalities in creating new affordances was also a key finding in Lu et al.'s StreamSketch, which allows viewers to offer their drawing suggestions through interaction on the stream's video player \cite{lu2021streamsketch}. Chung et al. developed VisPoll, a polling system using visual communication that can be aggregated by the streamer and summarized for the viewers, and found that the visual input led to increased expressiveness for viewers \cite{chung2021}. From all these works, we learn that visual input extends the limited expressiveness in text-based APGs \cite{lessel2017}, which our work also investigates and designs for. This is corroborated through research that has shown that interactive visual-based inputs can produce spontaneous and playful interactions \cite{Kazi2014} and increased reflection and understanding \cite{lu2018streamwiki}. 

We also consider some commercial alternatives that have also been developed to help integrate novel interactions into specific games. Software systems like CrowdControl\footnote{\url{https://crowdcontrol.live/} [Last accessed: March 12, 2025]}, Streamer.bot\footnote{\url{https://streamer.bot/} [Last accessed: March 12, 2025]}, and Dixper\footnote{\url{https://dixper.gg/landing} [Last accessed: March 12, 2025]} offer unique integrations into games. Whereas these applications do offer direct spatial input through audience interaction, their impact on audience-streamer motivations and relationships has not been heavily explored in academic research. 

Overall, many of these prior works, e.g., StreamSketch \cite{lu2021streamsketch}, Vispoll \cite{chung2021}, and Lessel's Hearthstone work \cite{lessel2017expanding} take mouse events to \emph{visualize} the aggregated data --- inputs are used to communicate \emph{about} the stream, but it remains the onus of the streamer to decide how to use this information. We are the first to investigate using such inputs directly \emph{into} streamer application, providing the audience with direct influence over the streamed content and giving rise to some level of uncontrollability for the streamer. 

\section{Developing the Technology Probe --- Building VIBES}

\begin{figure*}[h]
\centering
\includegraphics[width=0.9\textwidth]{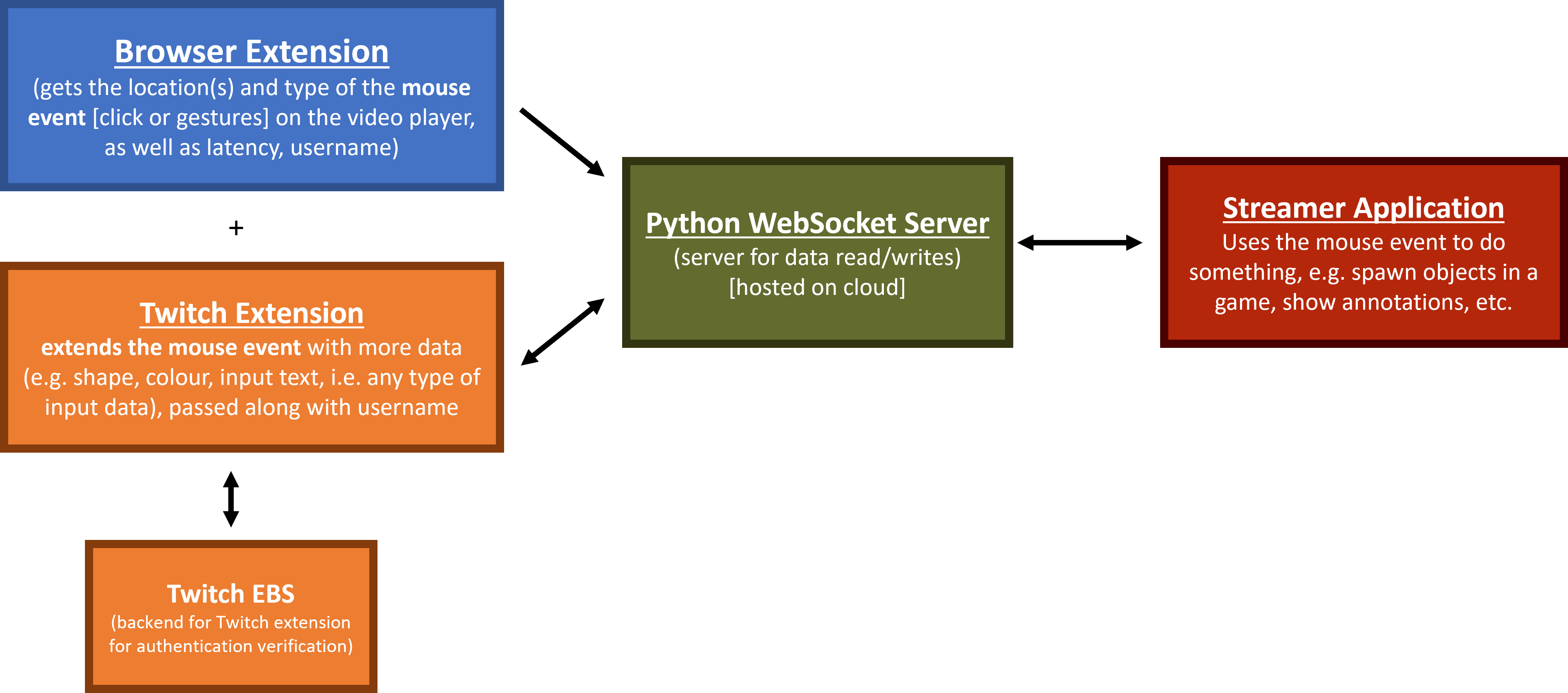}
\caption{Simplified architecture diagram of VIBES. The browser extension captures a viewer's mouse input; the Twitch extension provides added expressiveness to the spatial data. This data is passed to the websocket server, which transmits it to the streamer's application. The bi-directional nature of the architecture allows the application to send data back to influence the Twitch extension. }
\Description{A flowchart diagram outlining the architecture of the VIBES system. There are four nodes in the flowchart relating to the four components of the system. The browser extension and Twitch extension on the left are audience-facing, and the streamer application on the right is streamer-facing. Communication is facilitated through the websocket server in the middle.}
\label{fig:setup}
\end{figure*}

This section summarizes the implementation and features of VIBES. Although VIBES was used primarily as a means of exploring our research questions, there were several key design challenges that we came across, outlined here. 

\subsection{Design Decisions, Goals, and Objectives}

We motivate our decisions of platform and medium, which helped in focusing our exploratory probe. We chose Twitch as our platform for its readily-accessible and extensible API and its prominence in prior livestreaming literature. We chose to develop for desktop-based systems as a medium because of technical restrictions on extending the Twitch experience to mobile devices. While it is relatively easy to integrate 3rd-party features into web browser applications, it is much more difficult on a mobile application; at the time, the mobile version of Twitch also lacked several key features necessary for our desired interactive system. In developing a robust system to translate spatial interaction events, we focused on three primary goals:
\begin{itemize}
    \item \textbf{Functionality} --- developing a system that would be able to transmit real-time mouse events on the video player to the backend without error. This was crucial for ensuring reliable interaction in livestreaming environments, where synchronization is expected. 
    \item \textbf{Adaptability} --- building a system that could be extensible to a variety of applications to support the broad use cases in interactive livestreams, given its use as a probe \cite{techprobe}. 
    \item \textbf{Expressiveness} --- extending simple mouse events into complex actions, enhancing the depth and richness of interaction that viewers could use to impact the streamed content.     
\end{itemize}

\subsection{Architecture and Implementation}

The system architecture of VIBES consists of 4 main components (Figure 2).

\begin{itemize}
    \item A \textcolor{blue}{\textbf{browser extension}} running on the viewer's computer captures the type (either a click or a gesture) and coordinates of the mouse event, as well as the viewer's username and the stream latency.
    \item An optional \textcolor{orange}{\textbf{Twitch extension}} provides additional data packaged with the mouse event (e.g., a clicked shape) along with the associated username. This is hosted on the Twitch servers and can be activated by the streamer, varying depending on the streamed application.
    \item Both these components send data to the \textcolor{olive}{\textbf{websocket server}}, which is hosted in the cloud. 
    \item The server then passes the data to the \textcolor{red}{\textbf{streamer application}} running on the streamer's computer, which decodes the data to perform an action in the application (e.g., spawn an item at a selected coordinate).
\end{itemize}
    
The streamer's application can also send data back to the viewer’s client; this can be used to communicate changes to the Twitch extension. VIBES is our implementation of a system that translates spatial mouse events into input for livestreaming applications. It is a lightweight, flexible system that can adapt based on the specific streamed content and the desired use of inputs. We discuss more detailed information for each component in the following sections. 

\subsubsection{Collecting Mouse Events from Users}

VIBES collects mouse events from a viewer using a \textcolor{blue}{browser extension} that activates when viewing a Twitch stream. Spatial interaction events on the browser can be dimensionalized through clicking and movement, leading us to develop two main events as the vocabulary for our exploratory study --- (1) mouse clicks --- an instantaneous click on the video stream, and (2) mouse gestures --- a click-and-drag motion on the video stream. This extension retrieves data regarding 1) the type of mouse event performed and 2) its spatial information relative to the video player. To get the spatial location of the mouse click and motion, we localized and normalized the mouse position relative to the video player element, obtained through query selections on the DOM. To differentiate a gesture from a singular click, we set a timeout and motion threshold starting from when the user presses the mouse down. DOM manipulation was used to automatically retrieve the broadcast latency of the stream from the Twitch video stats panel (i.e., pipeline latency associated with streaming real-time content \cite{latency}). Once these values were obtained, they were used to generate a JSON object, which was then sent to the \textcolor{olive}{websocket} hosted on a cloud server. 

\subsubsection{Adding Flavour to Events through Twitch Extensions}

Having only mouse events as input is rather limiting; we used \textcolor{orange}{Twitch extensions} to generate flexibility and expressiveness by allowing users to augment their mouse events with contextual data. These extensions are inline frames (iframes) that can be added to a stream and are hosted on Twitch servers. Viewers can pass data through forms, button clicks, etc., following the usual suite of web interactions. This data, along with the username of the viewer, is collected and sent to the websocket and then to the \textcolor{red}{streamer's application}. The streamer application then stores this extension information using the viewer’s username as a key for a hashed key-value pair. Once a mouse event by the same user is read by the application, it retrieves the hashed value, and that user’s data can be used in the application logic to generate interesting outcomes beyond simply spatial coordinate data. For example, a viewer could first click on the overlay iframe to select an item and then spawn the item into the application at a specified location by then clicking on the video player. 

\subsubsection{Passing Data back from Applications} 

To adapt VIBES to diverse stream applications, we provided the ability for the streamer application to pass data back to the viewer (the Twitch extension) so that there could be bidirectional interaction --- this allows the extensions described in the prior section to adapt towards the streamed application. For example, there may be times in which the streamer applications need a change in the level of expressiveness, to lock out viewers from providing spatial interactions, and so forth. Thus, data can be passed back from the streamer application to the websocket and later propagated to the client Twitch extensions. 

\subsubsection{Dealing with Broadcast Latency}

One major challenge that immediately became apparent was that of broadcast latency, the inevitable delay in the stream as it is presented to viewers \cite{Zhang2015}. Even at ultra-low latency settings, there exists a lag between 0.2 and 2 seconds between the streamer's and viewers' perspective \cite{latency}; this exceeds the delay for interaction instantaneity that users may expect \cite{Dabrowski2001}. During this loading period, the viewer receives feedback in the form of a loading circle whose countdown animation time is equal to the broadcast latency obtained from the Twitch video stats panel. Through informal testing over a 5-minute stream, we found that this latency variable differs from the screen-to-screen latency by a mean of 233 ms (s.d. = 66 ms), which is adequate for our usage. The major advantage of using this scraped latency variable is that it can be automatically retrieved and updated for each viewer, eliminating the need for manual clock calibration of timestamps as in prior systems, e.g., Helpstone \cite{lessel2017expanding}. Although there were other sources of delay as well (e.g., web requests), this was largely negligible in comparison.

Another issue with broadcast delay involves matching user action with intent. When a user performs a mouse event, they perform it on the video player that is showing the application at a time in the past relative to the streamer. However, the input is received by the application near-instantaneously. If the application's appearance changes within the broadcast delay period (e.g., if the application’s camera moves), this will cause a dissonance in intent --- should the user mouse event be registered in regards to the current camera or the past camera of the application? In our system, we chose the latter, as we believed that the latter would more closely align with viewer intent, i.e. the viewer is interacting with the stream \emph{as they presently see it} rather than the streamer's live state. As the stream from the viewer's perspective happens in the past from the present stream, this implies we must keep a record of what happened in the past to reconcile this latency. In our applications, we considered that the most common aspect that may change in a single streamed application is the ``camera'' (for games, but more specifically the ``viewport''). 

To address this, we implemented a buffer - a time-shifted record of the application's visible interface (e.g. game camera position, scroll location, etc.). In our demo systems, we retained the past 10 seconds of data (given the assumption that latency would not surpass this period) and updated every 0.1 seconds. Thus, when the viewer triggers a mouse event, the system 1) notes the broadcast latency for that user (which is already logged from the browser extension), 2) retrieves the historical camera state from the buffer, and 3) applies the input relative to that camera state. Currently, the implementation of this optional camera buffer is the responsibility of the application developer; however, in the future, it could be abstracted directly into the system to simplify integration for applications with dynamic, moving interfaces. Ultimately, broadcast latency was a major identified issue in development --- viewers inevitably interact with a time-shifted perspective of the stream. However, we aimed to design for it rather than overlook the issue. 

\begin{figure*}[!htb]
\centering
\includegraphics[width=0.75\textwidth]{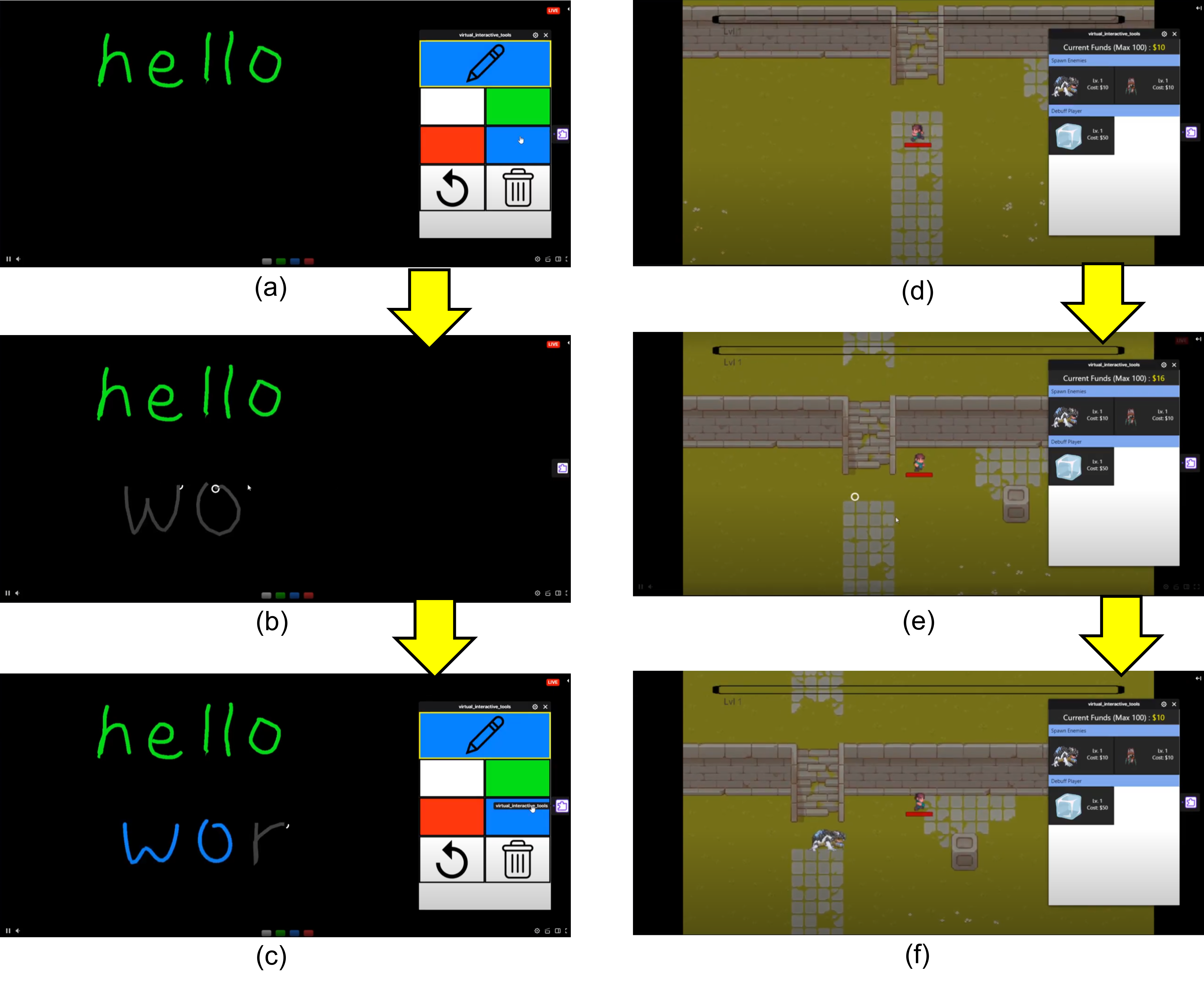}
\caption{The left is an application for drawing on a shared canvas. In \emph{(a)}, the viewers use the Twitch extension to select colour, undo, or delete. In \emph{(b)}, the user uses their mouse to draw the letters ``w'' and ``o''. Although small, note the spinner circles used to denote latency. In \emph{(c)}, the user's drawing has been passed to the Unity application, where it renders the line with the correct colour onto the application. The right is an interaction for spawning an enemy into a game. In \emph{(d)} viewers can select which enemy or debuff to apply. In \emph{(e)}, viewers click into the game where they want the enemy to spawn and in \emph{(f)}, the viewers see the enemy spawned. 
}
\Description{The first three subimages depict the flow for a shared canvas application. In the first picture, the streamer has drawn the word ``hello'', and the viewer is using the Twitch extension to select a colour (blue). In the second picture, the user draws the letters ``w'' and ``o'' on the screen, which appear as a translucent outline with a spinner countdown. In the third picture, the letters ``w'' and ``o'' have been rendered in the application. The latter three subimages depict the flow of an application in which viewers can spawn enemies into the streamer's game. The first picture depicts the Twitch extension, which has components corresponding to the enemy or debuff they might want to deploy. In the second picture, the viewer has selected an enemy and then clicked into the video player at the location they would like the enemy to spawn. In the final picture, the enemy spawns at the corresponding location in the game. 
}
\label{fig:Demo34}
\end{figure*}

\section{Demo Interactions}

With VIBES built, we brainstormed potential ways in which direct spatial input could be incorporated into stream applications, taking the perspective of the researchers as viewers. We built out these samples using VIBES, forming lightweight, illustrative interactions. One limitation of these demo interactions is that they were created and tested by the researchers in-house, with researchers acting as both the individual streamer and the viewers (i.e., using multiple accounts). This entails a very small number of viewers; scaling to a larger audience is important for future research.

\subsection{Sharing an Experience Together --- Participating With and Against the Streamer}
Viewers and streamers can participate in a single application using VIBES (Figure 3). In the first interaction, users can collaboratively draw together with the streamer on a shared canvas. Users can use the associated Twitch extension to select a colour before making the drawing on the screen. The coordinates are sent to the Unity backend, which processes them and renders the drawn lines with the appropriate colour. Viewers can furthermore use the Twitch extension to undo their prior input or to clear all of their inputs. 

In the second interaction, viewers can play against the streamer in a top-down 2D shooter game. In this game, viewers can spawn enemies by using the extension to click the desired enemy and then the video player at the desired spawn location. This game also illustrates a way in which the application data can be passed back --- when the streamer defeats a certain number of enemies, they level up, and this information is communicated to the Twitch extension. 

\begin{figure*}[h]
\centering
\includegraphics[width=0.75\textwidth]{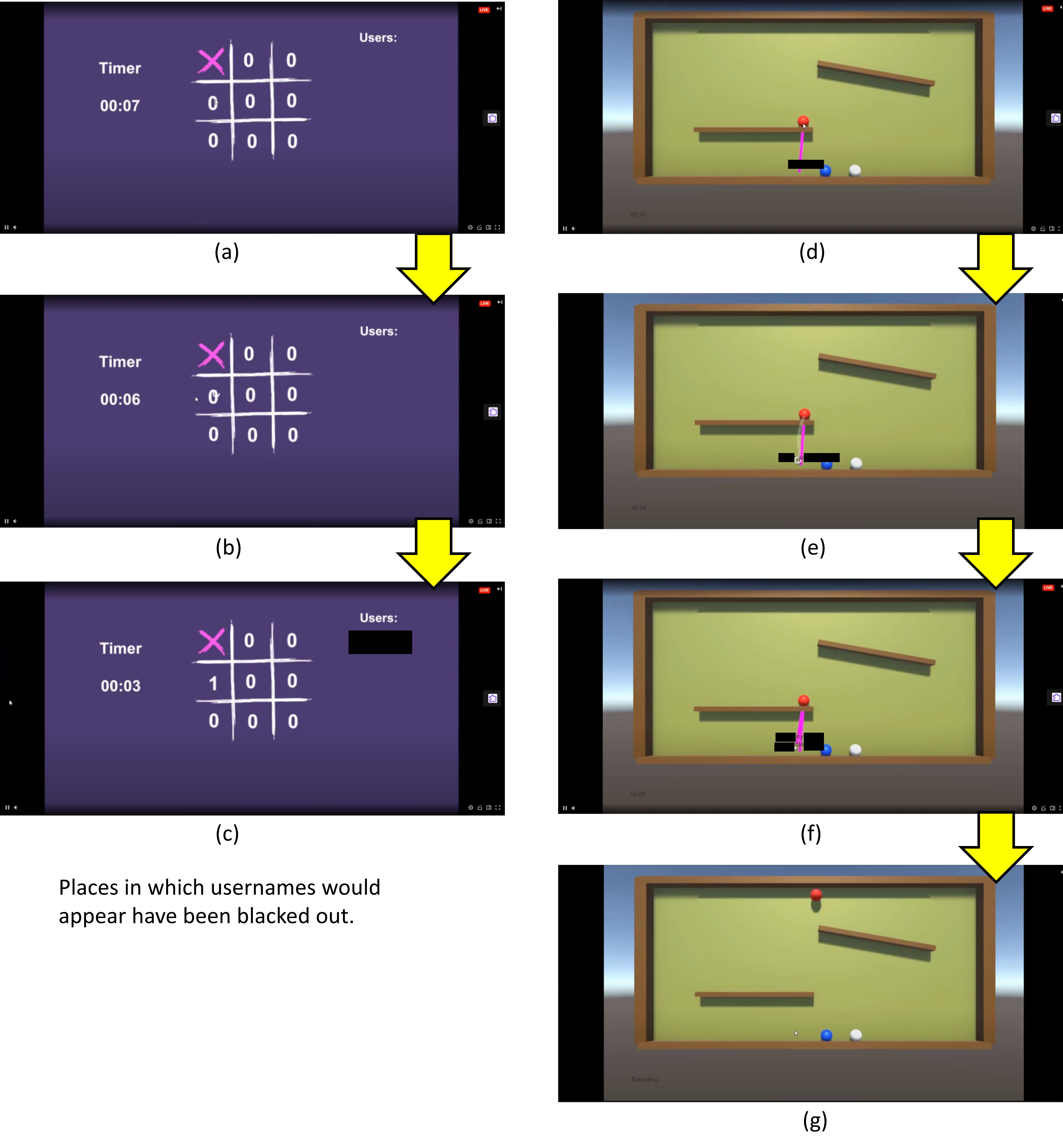}
\caption{The left interaction surveys clicks. \emph{(a)} illustrates the initial scene, where viewers are prompted to vote on a move. In \emph{(b)}, the viewer clicks to vote for their selection (the middle-left square); in \emph{(c)}, the number at this square increments and the username of the viewer is added to a list on the right of the stream. The right interaction surveys gestures. \emph{(d)} illustrates an initial scene, where a different viewer has already ``voted'' for a force. In \emph{(e)}, the viewer drags from the ball to add another force, which is then represented as a line in \emph{(f)}. Once the timer has completed, the average of the forces is applied, as seen in \emph{(g)}.
}
\Description{The first three subimages depict the flow for tic-tac-toe as applied using VIBES. In the first picture, there is a tic-tac-toe board and the streamer has made the first move --- an X in the upper left square. There is also a timer in which viewers can vote for a square. In the second image, the viewer makes their selection by clicking on the square they want (the middle left). In the final image, the number on the middle-left square increments from 0 to 1, and the viewer's username is added to a list on the right side of the stream. The latter four subimages depict the flow of an application in which viewers can employ forces on a ball in a contained environment. The first image contains an image of the application, where one of the balls already has a primed force represented by a line. In the second image, the viewer primes a force by drawing a line on the video player starting from one of the balls. In the third image, this drawn line has been used by the application to develop a line with the viewer's username, and in the fourth image, the forces are applied and the ball moves. 
}
\label{fig:Demo56}
\end{figure*}

\subsection{Polling the Audience, But with Spatial Information}

Streamers can use VIBES to survey their viewers through spatially-dependent polling, which we define as polls where responses are dependent on position or actions within the visual space, contrasting text-based polls (Figure 4).  The first interaction polls for spatial clicks in a game of tic-tac-toe. On the viewers’ turn, they can vote (click) on which square they would like to place their symbol in during the duration of a countdown. Each square's value increments upwards from zero as users click to vote. Under the hood, the spatial location of the click is projected as a coordinate onto the game plane (corresponding with a square if the user clicks correctly) and the user is registered to have voted. When the timer reaches zero, the application places the symbol onto the most-voted square.

The second interaction polls for spatial gestures in a ball-flicking game. Viewers can draw a line on the video screen by clicking on a ball on the screen and dragging their mouse to control the direction and force of the flick during a countdown timer. This becomes a “primer” for a force, which is illustrated as a line on the application screen. When the timer reaches zero, these forces can be aggregated and applied. On the backend, the application keeps track of the various mouse gestures and uses that information to determine which ball is affected and how the force should be applied, showing the result on the screen. 

\subsection{Interactive Controls and Annotations to Extend Stream Content}
VIBES allows viewers to interact with streamer software directly through gestural controls or annotations. For instance, we demonstrate how viewers can use mouse gestures to control the user’s music application. A user’s mouse movement is read by a backend Python application on the streamer’s computer and then translated through a simple gesture recognition system (Figure 5). The application then makes a call to the Spotify API, which then changes to the next or previous song, depending on the gesture. This integration pattern could potentially apply to any number of applications that the streamer uses, as long as they have a suitable API.

\begin{figure}[h]
\centering
\includegraphics[width=1.0\linewidth]{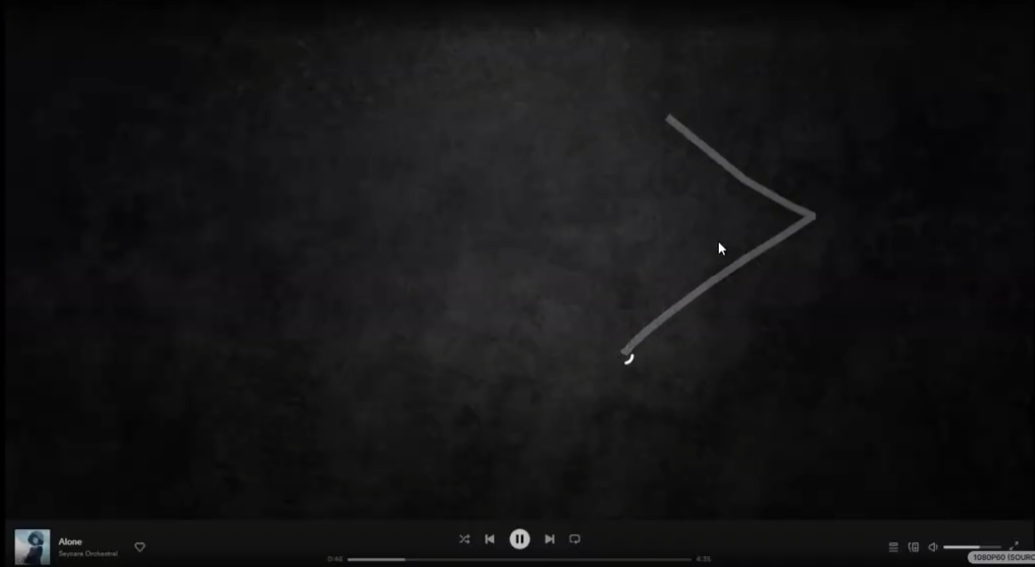}
\caption{VIBES can be used to input mouse gestures as control. For example, we developed a simple arrow gesture decoder that can be used to control the streamer's Spotify application.}
\Description{The viewer of the stream has drawn a ``>'' shape on the stream, indicating that they want the streamer's Spotify application to go to the next song.  
}
\label{fig:Demo1}
\end{figure}

\begin{figure}[h]
\centering
\includegraphics[width=1.0\linewidth]{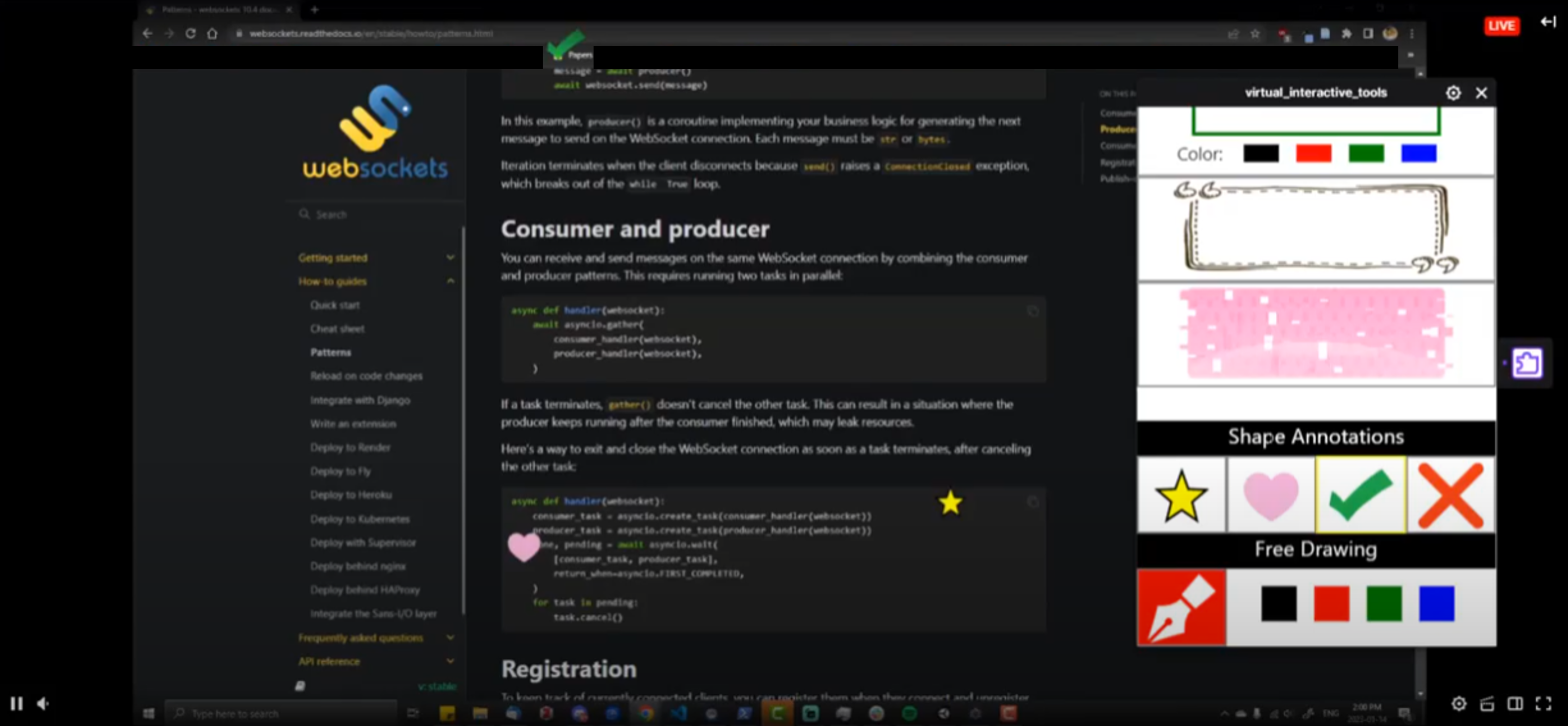}
\caption{VIBES can be used to annotate on the streamer's screen. On the stream, we see a heart, checkmark, and star icon that has been spawned in. This tool can also be used to spawn textboxes with viewer-specified text or to freehand draw by the viewers. }
\Description{The streamer is browsing the internet. The viewer has created annotations on the streamer's display (on top of their browser), including symbols of a checkmark, star, and heart. The Twitch extension is open, and using it, the viewer can also create a textbox with custom text or freehand draw on the streamer's display.
}
\label{fig:Demo2}
\end{figure}

Drawing inspiration from functions in online meeting software like Zoom but at a much broader audience scale, VIBES can facilitate annotations on the streamer’s screen (Figure 6). We showcase a suite of possible annotations for viewers --- text boxes, shapes, and freehand drawing. Users can select the desired annotation using the Twitch extension and click on the stream where they want the annotation to appear. An Electron web application renders the annotation onto the streamer's screen, which is seen by all viewers.

\section{Formative Exploration with Streamers}
With a prototype of VIBES and some illustrative interactions, we conducted an initial formative exploration with streamers to probe into how this technology might affect viewer interactions and motivations, how it could be implemented into practical streaming contexts, and what the possible future improvements could be, addressing \textbf{RQ1}. 

\subsection{Participant Recruitment}
We recruited participants who had a history of active streaming. Although VIBES was built as a Twitch integration, we were open to recruiting participants from a variety of streaming platforms, such as YouTube Live, Facebook Live, or even Zoom, as we were primarily focused on the technology-agnostic overarching concept of extending livestreams with mouse events rather than any specific technical implementation. We used a variety of recruitment channels, including personal connections, the institute’s paid studies listings, and posting on the /r/Twitch subreddit.

We were able to recruit 6 streamers for this study. The ages of our participants leaned young, ranging from 19 to 28 (mean: 23.5), with a gender distribution of 2 male, 3 female, and 1 non-binary. Although our sample size was rather small, we felt that it was sufficient for a formative exploration in terms of capturing meaning-richness and that it furthermore represented a diverse range of streaming practices --- most participants indicated that they streamed video games, but some also brought perspectives from educational streaming (e.g., online teaching and collaboration) or music-related streaming. A more detailed representation of the participants can be found in Table 1.

\begin{table*}
  \caption{Summary of Interview Participants}
  \label{tab:1}
  \renewcommand{\arraystretch}{1.2}
  \begin{tabular}{M{0.5cm}M{1cm}M{1cm}M{2cm}M{2cm}M{3cm}M{2.5cm}} 
    \toprule
    \textbf{ID} & \textbf{Age} & \textbf{Gender} & \textbf{Avg. Viewers} & \textbf{Hours Streamed / Week} & \textbf{Streaming Platforms} & \textbf{Primary Streaming Content} \\
    \midrule
    P1 & 22 & F & 10-50 & 5-10 & Twitch & Games \\
     \midrule
    P2 & 21 & NB & 0-10 & 10+ & Twitch, Tiktok & Games \\
     \midrule
    P3 & 28 & F & 0-10 & 0-2 & Zoom, Twitch, Discord, Teams & Games, Education \\
     \midrule
    P4 & 25 & M & 0-10 & 0-2 & Twitch, YouTube Live & Games, Music \\
     \midrule
    P5 & 26 & F & 200+ & 0-2 & YouTube Live, Facebook Live, Zoom & Music, Performing Arts \\
    \midrule
    P6 & 19 & M & 0-10 & 0-2 & YouTube Live, Zoom, Google Meet & Education \\
    \bottomrule
  \end{tabular}
  \Description{List of information about participants, including age, gender, average viewers per stream, hours streamed per week, streaming platforms, and primary streaming content}
\end{table*}

\subsection{Study Protocol}
Before the study, participants were asked to review and sign a consent form, outlining data collection and ethics approval; audio recording was collected with consent. The study was done through remote audio calls over Zoom and was split into two sections. The introductory part was a semi-structured interview focused on the streamer’s stream activities and motivations, from which we explored what, how, and why participants stream, as well as their interactions with their viewers. Examples of questions asked in this section were \emph{“What are ways that you currently engage with the audience?”} and \emph{“What are ways in which the audience can communicate with you?”}. These questions aimed to illuminate the streamers' present motivations for streaming and help highlight possible ways VIBES can impact such motivations. 

The second part of the study focused on addressing \textbf{RQ1} through our technology probe. To get participants familiar with VIBES, we presented demonstrative videos from the applications described prior, plus an additional one to show the effects of latency. Although videos do not fully capture the impression of the system compared to a full demo, they were convenient for a formative exploration --- setting up the actual interactions takes significantly more time and effort on the participant's end. While watching the videos, we asked the streamers to freely talk about their thoughts before later delving into a discussion focusing on how direct spatial input might be integrated for unique streamer interactions, as well as the limitations and challenges presented by the idea. Studies generally took around an hour, and participants were compensated with \$16 CAD for their participation. 

\subsection{Data Analysis}
The collected qualitative data was examined and analyzed through a content analysis approach \cite{eloQualitativeContentAnalysis2008, hsiehThreeApproachesQualitative2005}. After familiarizing ourselves with the data, the qualitative texts were then coded by the primary researcher \cite{saldana_coding_2021}, with feedback and discussion from the other researchers to attenuate single-coder bias; nonetheless, we note that coder bias could be a possible limitation of our study. These initial codes represented the base, semantic content of the textual data. Afterward, the codes were iteratively refined and related codes were grouped into broader categories. The final codebook consisted of 24 codes and 6 categories; some examples of the final codes were \emph{``Motivations for Interactivity''} and \emph{``Potential Applications''}; some of the final categories were \emph{``Engagement and Interactivity''} and \emph{``Experience and Intentions''}. The process of content analysis helped us garner a holistic understanding of how streamers perceive and might integrate direct spatial input into their practices, helping us answer \textbf{RQ1}.

\section{Findings From the Formative Exploration}

\subsection{Streamer Perspective}
\emph{\textbf{RQ1} --- How does direct spatial input affect the streamer’s experience, and what challenges and opportunities emerge for streamers?}

\subsubsection{Extending Audience Engagement, Which is Important for Streamers}
All streamers were overall positive and receptive to VIBES as a way of extending audience interaction. Although they indicated that the level of engagement might differ depending on the application they were streaming, all participants emphasized the importance of audience engagement and interaction as a reason and motivation for streaming, agreeing with prior research \cite{Taylor+2019, li2020, wohn2020, lu2019}. 

Although engagement was usually indicated to be done through chat, streamers pondered on potential applications of direct spatial input in their livestreamed content. Through their responses, we identified two main distinguishing features of such input that allow for potentially novel interactions --- \textbf{visual spatial encoding} --- the ability to encode visual information, and \textbf{continuous inputs} --- the ability to provide gestural or movement-based input. Regarding the former, P2 mentioned that this tool can allow the audience to help them in games by providing visual information in a game --- \emph{“I got lost a lot... I would love this. People have been pointing where I have to go”}. This is expanded by P1 -- \emph{“you can actually draw the direction that you want the player to go to”}, which P3 expands as \emph{“360-degree movement; there's so many possibilities of where someone could draw the gesture”}. 

Altogether, the streamers indicated that VIBES could promote a more collaborative and interactive environment between streamer and audience, e.g., \emph{“I think this will just add more options for viewers and streamers [to control] how much interactivity they want”} (P4). P1 hypothesizes a collaborative drawing tool that could provide \emph{“more freedom to the viewer”}. VIBES could also create a more continuous and seamless stream experience. The streamers indicated that they had to actively keep up with chat, but \emph{“if you're able to kind of simulate that with gestures instead on a screen... [it] actually reduces how often you, as a streamer, need to look at a message”} (P3), and, regarding the polling interactions, \emph{“this one is actually pretty cool for like choice-based games ... because sending up a poll, it just takes too long”} (P2). 

\subsubsection{Impacting the Balance between Streamer and Viewer Intention and Control}
Streamers noted that direct spatial input could complicate their intentional control over the streamed content. The content of a livestream is typically under the full authority of the streamer, yet allowing users to provide direct input alters the balance of agency. As such, the stream is then influenced by viewer intention, and streamers discussed the challenge of matching viewer intention and desires to the streamer's desired experience. For example, P1 mentioned that if viewers continually use the system to spawn enemies behind a streamer, \emph{“[the streamer] won't even know it... I think, after a bit of the streamer getting killed every two seconds spawning in and dying, they're probably gonna get tired of it”}, highlighting a possible mismatch between viewer and streamer desire. P3 stressed the importance of matching viewer intention and streamer expectations, suggesting increased streamer control to help manage this dynamic, e.g., \emph{“You can add a setting from the streamer because then it's based on the streamer's comfort level”} (P3).

Streamers expressed concerns over matching intentions given the inevitable latency in livestreaming. As the streamer content is slightly ahead of what the viewer sees, scenarios in which the intentions might mismatch may be common, especially if the broadcast camera shifts during that time. Preliminary ideas were suggested to address this issue, e.g., \emph{“There should be some kind of notification [if the viewer makes an annotation]”} (P3) and \emph{“I think different sound effects, different colours”} (P4), recognizing that \emph{“any application where the world moves around is going to be difficult”} (P4).

\subsubsection{Raising Possible Issues Related to Usage and Moderation at Scale}
More speculatively, streamers noted that practical scaling to a large number of participants could pose a challenge --- if everyone could pass direct spatial input at all times, this would likely inevitably lead to \emph{“a really big mess”} (P2). One proposed solution to deal with a massive amount of input data was filtering by who can provide inputs, e.g., P2 indicated that you can make the tool available for \emph{“only subscribers, or only VIPs, or only mods”}. Tying interactions to points and rewards on the livestreaming platform was also a popular suggestion, e.g., \emph{“you can leave a note on my screen for points or bits”} (P2), as was adding a timeout or cooldown for interactions, e.g., \emph{“it goes into a global cooldown”} (P2).

Moderation at scale was another speculated challenge. With more viewers comes more trolls --- unwanted agents that disrupt the stream or harass the streamer, and direct spatial input adds another way of accomplishing this. One way of dealing with these unwanted agents, in addition to existing methods like timeout or banning\footnote{\url{https://help.twitch.tv/s/article/how-to-manage-harassment-in-chat?language=en_US} [Last accessed: March 12, 2025]}, may be through input moderation at scale. P3 brought up a scenario in which a streamer allows free drawing on their application, and a viewer chooses to draw something inappropriate. P3, P4, and P6 indicated that detection algorithms could be used to filter out these inputs, \emph{“train something to detect an image of like a swastika and then remove that”} (P3). Reactive mechanisms for moderation (such as individual bans or restricting the tool to specific individuals) could also be made available by tying interactions to usernames. We highlight that issues regarding scale were more hypothetical in this context, as our work focussed on smaller, controlled audiences. Still, these considerations inform the importance of robustness and required safeguards for a real-world deployment. 

\section{Preliminary Application Investigation}

The prior formative exploration focused on the abstract potential of direct spatial input on livestreamed content; to attain more practical insights, we investigated its deployment under controlled livestreaming environments. We recruited two participants from the first exploration (P1 and P4), and based on their input and feedback, we developed an individualized application for each of them using VIBES. We then ran these applications in 2 livestreams with recruited viewers. Our main goal of this investigation was to address \textbf{RQ2} --- looking to investigate how extending actual livestreams with mouse events might affect the \emph{viewers} in terms of metrics of motivation, including participation, engagement, and enjoyment; but we also extended our findings on \textbf{RQ1} through assessing the experience of the involved streamers.

\subsection{Application Development}
A brief initial meeting (around 15-20 minutes) was conducted to collect information on the type of application they envisioned for facilitating spatial input. The streamers were kept informed regarding progress and provided feedback regarding potential changes. P1 described the development of a mod to an existing game (Terraria) in which viewers could collaboratively interact with the streamer; P4 proposed a tower-defence-based game in which viewers could help the streamer through spawning traps and picking upgrades. We developed these applications, respectively known as \emph{Terraria Interaction Mod} and \emph{Storm the Village}, and describe them in brief below (detailed information can be found in the supplemental material). Streamers were compensated \$24 CAD for their help in designing the application. 

\subsubsection{Terraria Interaction Mod}
Terraria is a bestselling sandbox-based adventure game in which a player can explore, build, interact with NPCs, etc., in a procedurally-generated world \cite{terraria}. Our mod (Figure 1) does not change the overarching mechanics of the game but provides viewers with three points of interaction:

\begin{itemize}
    \item \textbf{Spawning Items and Enemies} --- Audience members can select an item or enemy from the Twitch extension and spawn this selection at a cursor location within the game (there is a minimum distance at which enemies must be spawned). To limit the number of spawned objects, there is a funding system that provides funds at a rate inversely proportional to the number of viewers. 
    \item \textbf{Writing a Message} --- Audience members can write a message in the Twitch extension in an input field and then click into the game to spawn that message into the game for a few seconds at the clicked location. 
    \item \textbf{Voting on NPC appearance} --- During the night phases of the game, audience members can vote for which NPC they want to arrive the next day by clicking on the voting icons on the screen. 
\end{itemize}

\subsubsection{Storm the Village}
Storm the Village is a roguelike tower-defence game in which the player must repel waves of enemies from a castle by shooting them with their cursor, inspired by a similar game ``Storm the House'' \cite{stormthehouse}. Storm the Village was developed in Unity by the research team. The game operates on a day system --- during the day, the viewer shoots enemies that gradually become stronger. After each day, there is a vote for an upgrade --- these can provide new guns, increased gun damage, etc. To win, the player must repel 8 days' worth of enemies before the enemies manage to deplete the health of the castle. There are three points of interaction for the audience members (Figure 7):

\begin{itemize}
    \item \textbf{Spawning Traps} --- Viewers can select a trap from the Twitch extension and click onto the video player to spawn it at the clicked location. Traps deal damage to enemies and apply status effects, and more traps are unlocked as the game progresses. To limit the traps spawned, there is a funding system --- users must spend in-game currency (that increments at a constant rate) to build a trap. 
    \item \textbf{Voting on Upgrades} --- At the end of each day, audience members can vote for which upgrade they want the streamer to have by clicking the option directly on the stream. 
    \item \textbf{Freehand Drawing} --- There is a sequence in the post-game in which the audience members can draw something on the video screen. 
\end{itemize}

\subsection{Participant Recruitment}
We had initially intended for the streamer involved in the development to stream the corresponding application in the live session deployment. However, P1, who was involved in the development process of the Terraria mod, was ultimately unable to attend the livestream session. Thus, we recruited a new streamer (P7 --- gender: female, age: 20) who had prior experience with both livestreaming and the Terraria game; as such, we found them suitable for streaming the \emph{Terraria Interaction Mod}. For the other stream (\emph{Storm the Village}), P4 was able to attend and stream the application. 

For P4’s stream (Session 1), we were able to recruit 16 viewers (average age: 25.3, std: 3.91, ranging from 19 to 33; 13M, 3F) and for P7’s stream (Session 2) we were able to recruit 18 viewers (average age: 23.7, std: 4.35, ranging from 19 to 38; 14M, 3F, 1 Preferred not to report). For standardization, the participating viewers were not recruited as prior viewers of the assigned streamer; furthermore, there was no overlap between recruited viewers across the two streams. The eligibility criteria for recruitment were rather general in terms of previous livestreaming experience, aiming to develop broader results regarding the use of direct spatial input for those at any point on the spectrum of livestreaming familiarity. Participants were recruited through a mix of voluntary responses from our institute's paid studies listings and convenience sampling. 

Before the study, streamers and viewers alike were asked to read and sign a consent form relating to data collection and usage. Each of the recruited viewers was asked to install the browser extension and to provide their Twitch username for access to the Twitch extension. In addition, we provided each of the viewers with a document with information regarding the stream that they would be watching, outlining the game and the various interactions they could engage in. Viewers will henceforth be denoted by combining the participant ID of the streamer they watched with a viewer ID, e.g., P4V3 will refer to viewer ID 3 that watched P4’s stream. 

\subsection{Study Protocol}
The livestreams took place on the Twitch platform. Streamers were asked to stream on a researcher-created account, providing us with control over the distribution of the Twitch extension and the viewership. During the livestream session, the researchers were largely hands-off, letting the streamer take control over their style of streaming, their interactions with viewers, and the overall atmosphere. The streamers played the customized games, and the viewers were able to interact with the streamer using direct spatial input as well as normally through chat. We recorded the stream locally and saved a private video-on-demand (VOD) of the stream and a transcription of the messages sent in chat. Overall, the stream sessions took about 45 minutes. 

After the stream, viewers were asked to answer a survey regarding their experience. This survey combined two scales --- we first employed the questions from Chung et al.’s study to evaluate implementation factors such as the usefulness and helpfulness of the tool in achieving interaction with the streamer and other viewers (including open-ended questions for expanding on their rationales) \cite{chung2021}, then asked participants to fill out the \emph{Audience Experience Questionnaire (AEQ)} (a questionnaire developed to assess audience members in social video gaming sessions) to evaluate on the subscales of enjoyment, mood, game engagement, social engagement, and participation within the study \cite{downs2013}. We also performed a brief exit study with the streamer. Streamers were first asked to fill out the \emph{Game-Specific Attribution Questionnaire (GSAQ)} to gauge their attribution of the events of the game; we only used internality and controllability subscales to assess how in control the streamer felt over their experience \cite{depping2017}. We briefly interviewed the two streamers about their experience using VIBES, what they thought could be improved, and how it affected their perceived engagement with viewers. Viewers and streamers were compensated \$24 for their participation in the livestreaming session.

\begin{figure}[h]
\centering
\includegraphics[width=0.9\linewidth]{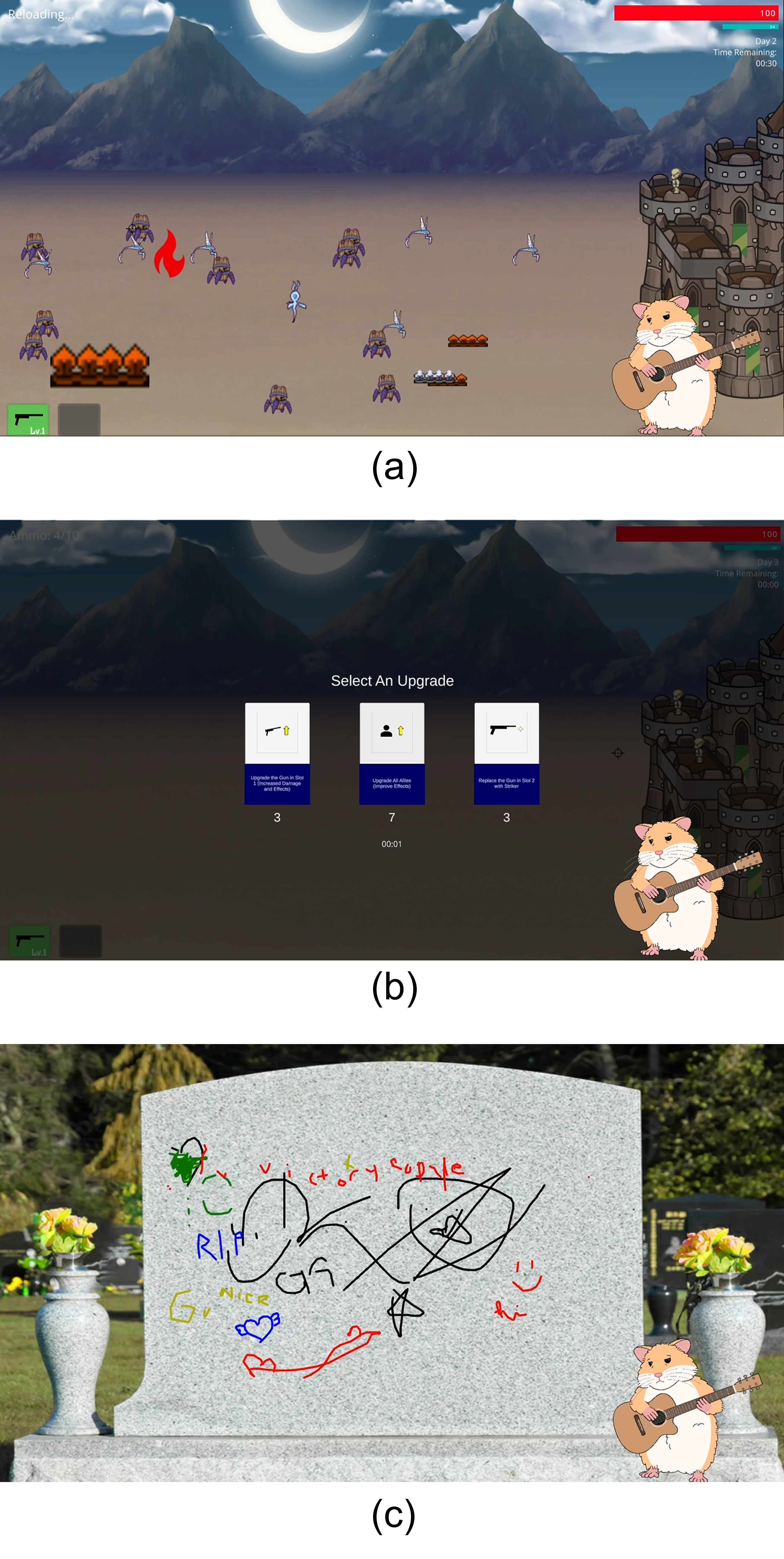}
\caption{
Images from the livestream of P4's study (Session 1) involving the game \emph{``Storm the Village''}. In \emph{(a)}, we see that viewers have spawned in traps to help the streamer repel enemies. In \emph{(b)}, viewers vote on the upgrade that they want. In \emph{(c)}, viewers have a chance to sign off or write something on a memorial after they have successfully beaten the game. 
}
\Description{
Images for P4's livestream (Session 1). The first subimage shows the game Storm the Village --- enemies are moving from the left towards a castle on the right and there are traps on the ground (spawned by viewers) to help impede and damage them. The second subimage shows the voting screen --- there are three buttons on the stream corresponding to specific upgrades, viewers can click on them on the video player to increment the votes. The final subimage shows the memorial that viewers can draw on after the game has completed. Viewers drew a variety of different shapes and symbols, such as a heart, a smiley face, etc. 
}
\label{fig:study1}
\end{figure}

\subsection{Data Analysis}
Our data consisted of both quantitative and qualitative data, which we analyzed using a mixed-methods content analysis approach \cite{eloQualitativeContentAnalysis2008}. The quantitative data (Likert-scale questions) was investigated through exploratory data analysis processes. Due to the exploratory nature of the study and the uncontrolled nature of Twitch streams, we refrain from presenting formal statistical tests \cite{vornhagen2020} and use them more as illustrative values. The graphs are presented separately for each study session due to the uniqueness of each streamed game and its associated interactions. 

For the qualitative data, we first performed initial open coding on the questionnaire responses to understand the impact of direct spatial input on viewers \cite{saldana_coding_2021}. We then refined, iterated, and grouped these codes into broader categories. Although the qualitative data was coded by the primary researcher, the analysis underwent discussion and feedback from the other researchers. For the exit interviews with streamers, we also underwent a similar coding process. In both cases, we took a largely deductive approach, contextualized by our formative exploration findings and background research. Overall, we developed 9 total codes (e.g., \emph{``Latency''}, \emph{``Agency in the Game''}) spanning 4 broader categories (\emph{``Challenges''}, \emph{``Engagement and Agency''}, \emph{``Enjoyment and Helpfulness''}, \emph{``Interaction with Agents''}) that tie towards both streamer and viewer experience, we developed into themes. 

We integrated the exploratory analysis of the quantitative data with the qualitative data to highlight similarities and potential disparities in the interpreted findings. Since the data sources were generally aligned in terms of their conclusions, quantitative evidence largely supports our qualitative findings. Our results primarily focus on the experience of the viewers due to the larger sample size, but we do provide brief deductive findings based on the streamers' experiences as well, extending our formative exploration findings in practice.  

\section{Findings From the Application Investigation}

\subsection{Audience Perspective}

\emph{\textbf{RQ2} --- How do viewers perceive and engage with direct spatial input, and how does it influence their interactions within the stream?}

\begin{figure*}[h]
\centering
\includegraphics[width=0.9\textwidth]{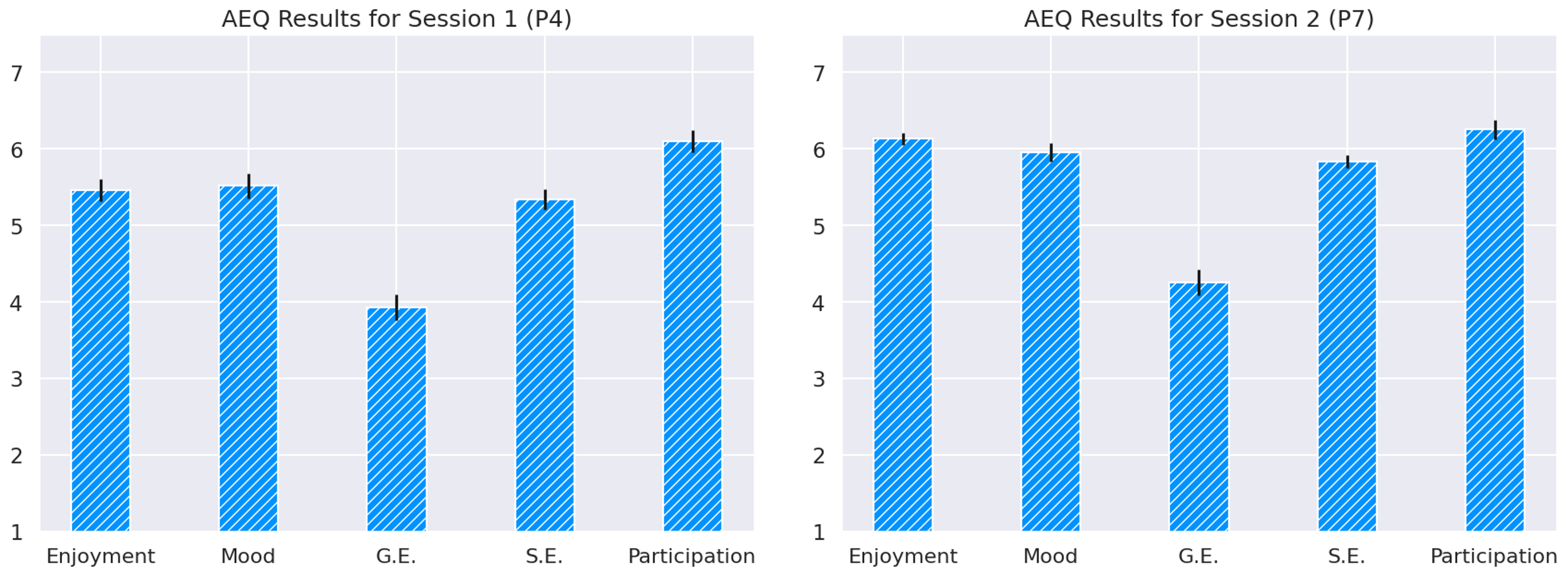}
\caption{
Mean AEQ values with 95\% confidence error bars separated by stream session, from 1 (strongly disagree) to 7 (strongly agree). G.E. and S.E. are abbreviations of game engagement and social engagement, respectively.
}
\Description{
Bar charts for the AEQ values grouped by subscale. Aspects of enjoyment, mood, social engagement and participation scored highly for both streams, but game enjoyment scored relatively lower. 
}
\label{fig:aeq}
\end{figure*}

\subsubsection{Generating Positive Feelings of Enjoyment and Engagement}
Direct spatial input was generally well-received by viewers from both an implementation and conceptual standpoint, evidenced by the responses to the questionnaire questions (Figures 9 and 10) and viewers’ open-ended responses. From the AEQ (Figure 8), participants scored highly on aspects of enjoyment (Session 1: mean = 5.5, Session 2: mean = 6.1), and mood (Session 1: mean = 5.5, Session 2: mean = 6.0). These metrics were supported through participants’ responses; for instance, \emph{``I felt like I was playing a game with friends which made my experience of the stream very fun''} (P4V5) and \emph{``it is a fun way to have creative interactions between the viewers and streamers''} (P7V11). Furthermore, almost all participants agreed that they would like to see other streamers use such tools within livestreaming (Q10 for Figures 9 and 10), \emph{“it would be amazing to see it integrated into a variety of games to change the way streaming is conducted and bring the viewer closer to the streamer”} (P4V14).


The use of VIBES to pass input positively affected viewer engagement and involvement for all participants, with every participant agreeing or strongly agreeing with the statement that they felt more engaged with the stream using the tool (Q3 for Figures 9 and 10). The responses to the AEQ support this sentiment, as the metrics of Social Engagement (Session 1: mean = 5.3, Session 2: mean = 5.8) and Participation (Session 1: mean = 6.1, Session 2: mean = 6.3) both scored highly across both study sessions. The active participation of the viewers within this stream was a contributing factor towards their enjoyment and engagement, e.g., \emph{“it helped me participate and actually play the game as well. So like I was able to have [a] direct effect on the game”} (P4V10) as was the real-time visual nature of the tool --- \emph{``I could see first hand the implications of my actions on the game ... it was really really interesting to see in real-time how my actions were affecting the game''} (P7V12). In particular, the idea of essentially co-playing a game in a significant way led people to feel that their actions were impactful --- \emph{“It felt like I was actually impacting the gameplay”} (P4V13). This level of engagement went beyond the traditional chat interaction --- \emph{“the interactive tools helped me engage more in the overall experience than Twitch chat”} (P4V5) --- perhaps due to the visual nature of seeing the results of one’s interactions, e.g., \emph{“being able to see the result of my interaction made it more interesting”} (P4V1). Overall, VIBES helped provide an increased degree of participatory agency for viewers by allowing them to influence the game.

The metric of Game Engagement scored relatively lower compared to the others (Session 1: mean = 3.9, Session 2: mean = 4.3). A hypothesis regarding the lower-scored metric may be related to the nature of streaming as a passive entertainment source in general, where many may have it open ``in the background'' as passive lurkers \cite{wyndow2022subculture, edelmann2017lurkers}. This relatively lower game engagement metric from the AEQ has been observed in other studies regarding livestreaming interactions \cite{robinson2022}, and could suggest that the engagement felt by the audience is more centred around the presence of others rather than immersion within the game itself \cite{downs2013}. 

\begin{figure*}[h]
\centering
\includegraphics[width=0.9\textwidth]{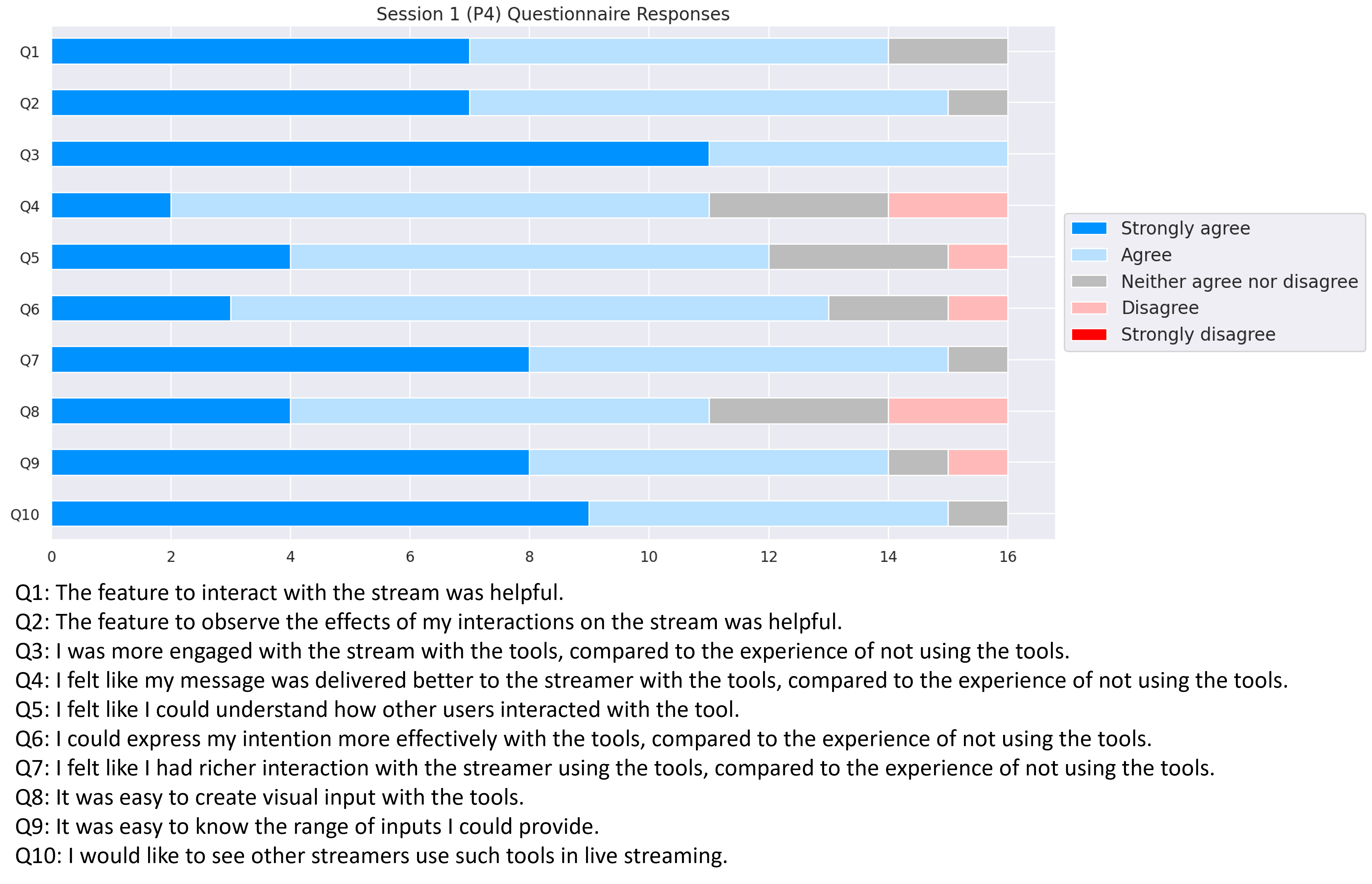}
\caption{
Raw distribution of questionnaire responses for P4's stream (Session 1)
}
\Description{
Stacked bar charts (for Session 1) showing the responses to the 10 questionnaire questions ranging from Strongly Agree to Strongly Disagree. These questions include statements such as ``The feature to interact with the stream was helpful'' and ``I would like to see other streamers use such tools in livestreaming''. The vast majority of responses lean towards Strongly Agree and Agree. No participant indicated Strongly Disagree on any of the questions. 
}
\label{fig:chung1}
\end{figure*}

\begin{figure*}[h]
\centering
\includegraphics[width=0.9\textwidth]{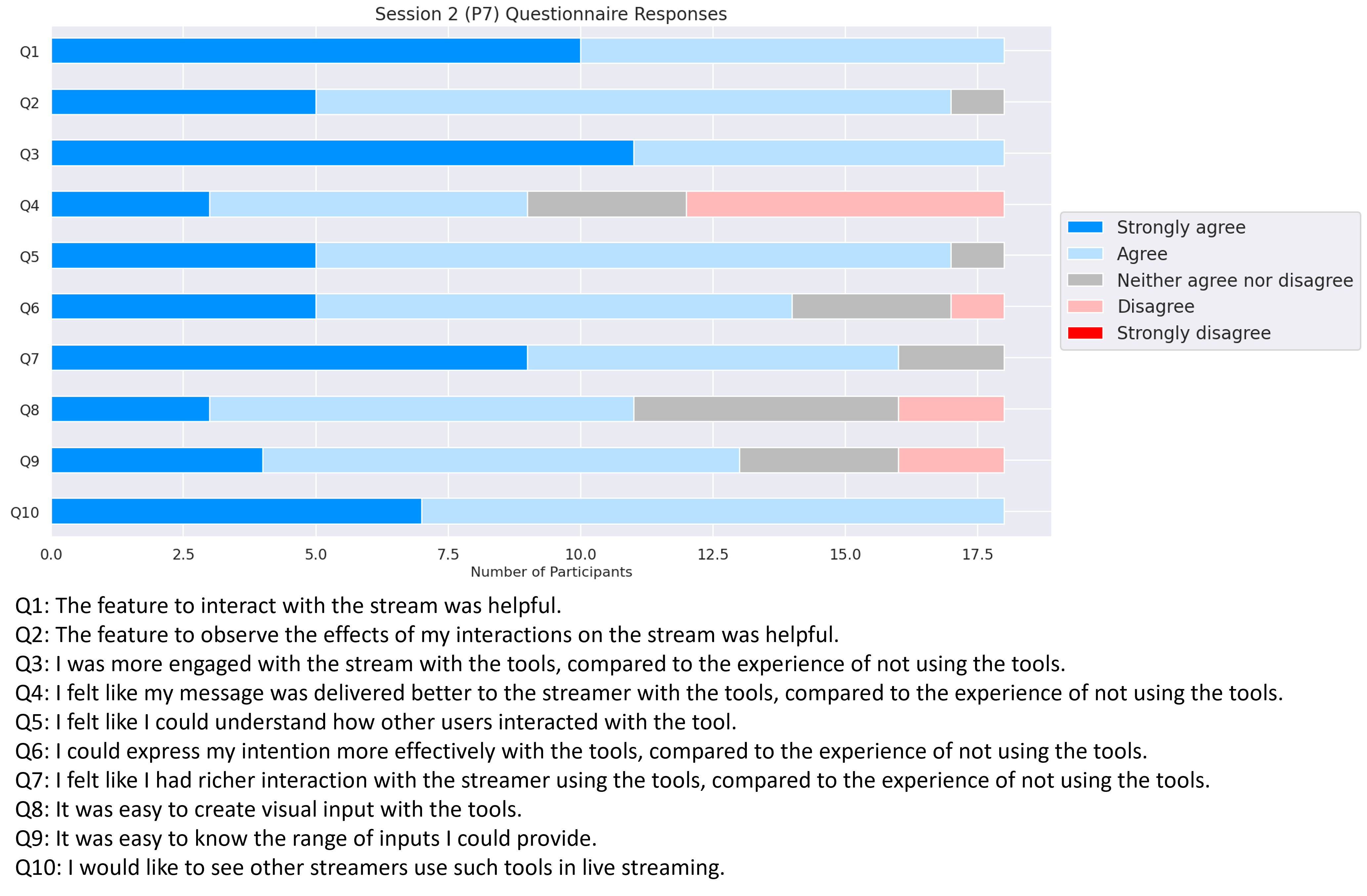}
\caption{
Raw distribution of questionnaire responses for P7's stream (Session 2)
}
\Description{
Stacked bar charts (for Session 2) showing the responses to the 10 questionnaire questions ranging from Strongly Agree to Strongly Disagree. These questions include statements such as ``The feature to interact with the stream was helpful'' and ``I would like to see other streamers use such tools in livestreaming''. The vast majority of responses lean towards Strongly Agree and Agree. No participant indicated Strongly Disagree on any of the questions. 
}
\label{fig:chung2}
\end{figure*}

\subsubsection{Creating Novel Forms of Audience-Streamer and Audience-Audience Interaction}

Interaction between streamer and viewer and between viewers themselves form important components of the livestreaming experience \cite{li2020}. Participants indicated that engagement brought forth from interaction induced increased forms of connection with the streamer, e.g., \emph{“it is a 2-way interaction unlike conventional streams which are 1-way”} (P4V12). P7V18 described it as a \emph{``more intimate interaction than just watching''}. Participants felt that the streamer interacted with the chat more because of the direct input because the direct impact of the viewers incentivized communication --- \emph{“as we were affecting the gameplay they would comment and interact with us more based on that”} (P4V14). 

Viewers also interacted with each other in interesting ways. The expressiveness of spatial input allowed for avenues of visual collaboration, e.g., \emph{“I could see where other users spawned traps and vote on the upgrade we needed”} (P4V11), although the intent was somewhat murky at times --- \emph{“harder to understand intent since I didn't know when they would place traps compared to me”} (P4V12). An interesting development was the use of chat for planning and coordination; which also led to an increase in feelings of camaraderie --- \emph{``being able to coordinate with and chat live with other participants. We could strategize together on how to attack (enemies) and also how to help''} (P7V12). Viewer coordination was observed on both streams. In Session 1, viewers collaborated within chat to vote on augments and spawn upgraded traps; in Session 2, viewers expressed intention in chat to save up funds to give the streamer enough gold coins to get a helmet. These provide examples of the symbiotic relationship that direct spatial input could have with chat, with chat providing a place for audience discussion revolving around interactions afforded by spatial input.

\subsubsection{Challenges with Latency, Scalability, and Feelings of Individual Agency}

Viewers also presented challenges and suggested improvements for the system. One problem that was brought up was that of latency, e.g., \emph{“it was interesting though slightly limited by delay.”} (P4V12). Even though our system tries to design for delay by using visual feedback to show that inputs are being processed and camera buffers to align the input with the viewer's tune-shifted perspective of the stream, we highlight delay as still a salient problem. We try to circumvent it through design, but its existence still negatively impacts the experience. Delay is inevitable given present livestreaming technology; however, viewers stated that a way to compensate for delay would be to have increased feedback on their interactions. This could provide increased recognition of interaction --- \emph{“sometimes I had trouble telling if my inputs were the ones registering''} (P4V8) --- and understanding of interaction effects --- \emph{“I think if the feedback is more obvious... [right now] I am not quite sure whether my interaction is that helpful”} (P4V7). This suggests that increased transparency and feedback for viewers regarding their inputs could improve the overall experience in terms of individual agency. 

Some viewers also raised scalability concerns. Viewers noted that the significance of an interaction might be relatively reduced with large numbers of viewers, e.g., \emph{“I think with more people, the less impact each person has with the tools”} (P4V8) and that a large audience involves the large degree of potential interactions --- \emph{``It might be overwhelming if too many people can spawn things at once''} (P7V16). Given the limited and controlled number of audience members in our study, this still requires future research. 

\subsection{Streamer Experience}

\emph{\textbf{RQ1} --- How does direct spatial input affect the streamer’s experience, and what challenges and opportunities emerge for streamers?}

\begin{figure}[h]
\centering
\includegraphics[width=0.95\linewidth]{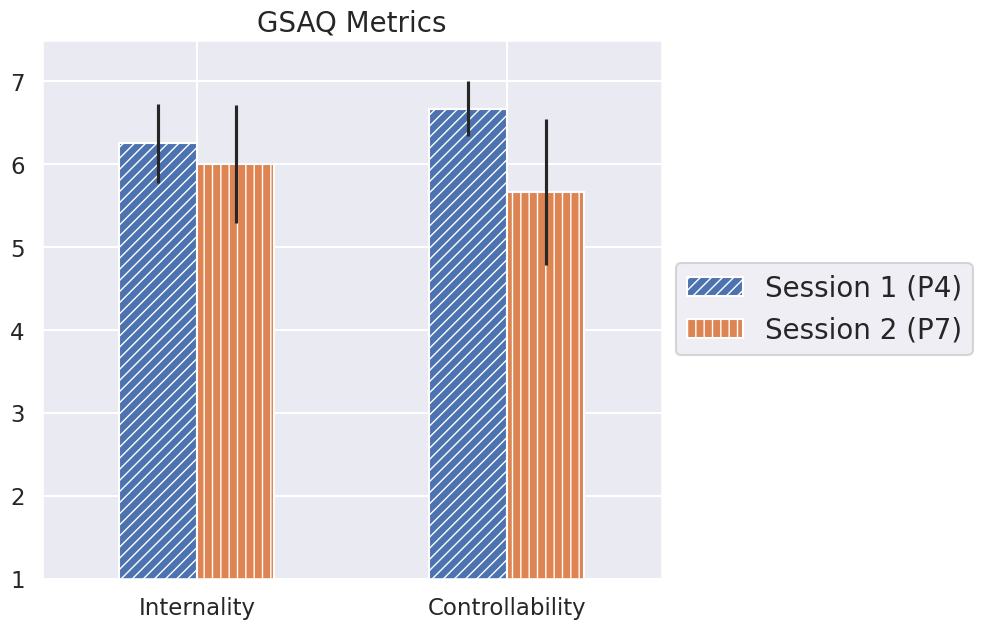}
\caption{
Mean GSAQ values with 95\% confidence error bars for the subscales of internality and controllability, from 1 (strongly disagree) to 7 (strongly agree). 
}
\Description{
Bar charts for the GSAQ values as reported by the streamer, grouped by subscales of internality and controllability. Both streamers indicated relatively high scores for both these metrics. 
}
\label{fig:gsaq}
\end{figure}

\subsubsection{Generating Enjoyment and Positive Engagement}

VIBES was similarly well-received by the streamers, e.g., \emph{``I think the game mechanic is cool and I got to see it in action''} (P4), \emph{“I think it was fun”} (P7). The streamers also noted that VIBES changed interactions between them and the viewers --- \emph{``because we’re playing the game together it does change how I interact. I probably was asking what the chat wants to do and such''} (P4). P4 also noticed how VIBES encouraged viewers to chat as well, as they would discuss which upgrade to get. Similarly, P7 indicated that during her stream, VIBES encouraged users to formulate a goal together --- \emph{“they had a goal that they wanted to accomplish which was to give me a bald cap [author note: bald cap = gold helmet]''}. Overall, much of the streamers' sentiments regarding engagement and interaction echoed those of the viewers, as well as those raised in the initial formative exploration. 

\subsubsection{Impacting Streamer Agency and Control}

We briefly explored understanding streamer agency and attribution while playing the games using VIBES, relating to the point of intentionality from the formative exploration. Both streamers scored rather highly on the metrics of internality (mean = 6.3 for P4 and mean = 6.0 for P7) and controllability (mean = 6.0 for P4 and mean = 5.7 for P7) on the GSAQ (Figure 11), implying that they believed themselves to have a high degree of internal control over their in-game performance. However, there were instances in which agency seemed to be mismatched in both games. P4 described that they thought \emph{“the player was too OP [overpowered]”} and that \emph{``for this game, I think it would be cooler if the power was shifted to the viewers''} (P4). In P7’s stream session, there was a moment where viewers collectively spawned many monsters at the same time, leaving the streamer unable to respond suitably and defend. Although this latter experience was chaotic and fun for the viewers, it shows that balancing streamer agency against viewer agency within the context of overall game balance becomes a key consideration, given the asymmetric nature of games involving audience participation. 

\section{Discussion}

We discuss the implications of our findings, contextualizing them around prior research into livestreaming, its motivations, and the overall experience for audiences and streamers. Furthermore, given that both studied applications were games, we discuss our findings through the lens of audience participation in game design. 

\subsection{Livestreaming Motivations and Experiences for Streamers and Audience}

Direct spatial input enhanced livestream engagement and participation, fostering closer bonds between the viewers and streamers, as well as among viewers themselves. The connections between viewers and the streamer tie towards the social motivator for livestream viewership \cite{lin2017, li2020}. Beyond just chat, spatial input generated a topic for communication with other audience members, which we interpret as strengthening the motivation of being alongside others in a common community \cite{lu2019, Taylor+2019}, and offering a sense of reward through direct impact \cite{hamilton2014}. By extending the range of interactions between streamer and viewer, spatial input fundamentally reinforces the motivations to watch and participate in streams \cite{li2020, lu2018}. Considering this from previously discussed theoretical frameworks for motivation, VIBES improved social engagement and interaction from a uses and gratifications perspective \cite{SJOBLOM2017985} and ties towards social benefit from a self-determination viewpoint \cite{zhaoDeterminantsLiveStreamers2018}. 

Both background research \cite{li2020, Taylor+2019, chen2021} and our formative evaluation revealed the importance of interaction and engagement for streamers. In our study, streamers generally noted how applications with VIBES encouraged inter-viewer discussion and increased social engagement relating to novel interactions. Tying into the theoretical framework, Young and Wiedenfeld indicated that this social engagement is important when relating to both self-determination and uses and gratification approaches regarding motivations for livestreaming \cite{youngMotivationAnalysisVideo2022}. Thus, our findings suggest that providing viewers with direct spatial input into the game can also positively affect the social motivators for streamers. Future work can consider extending the range of possibilities of such interactions, e.g. perhaps to leave persistent elements in the streamed content based on viewer input similar to messages in Dark Souls \cite{dooghanFantasiesAdequacyMythologies2025}. 

Taylor stated that, in livestreams, the audience and their interactions are integrated into the entire experience \cite{Taylor+2019}. This is particularly evident in our work, as the streamer and audience interact together. We observed a reciprocal push-and-pull feedback loop in the streams --- the streamer would interact with the audience, who would then use VIBES to interact with the streamer. This concept of reciprocity \cite{smith2013} allows for the co-creation of content \cite{smith2013} together with the streamer in a fundamentally different manner; direct spatial input allows the audience to more literally play together with the streamer. Spatially-mapped semantic information through mouse events provides an additional interaction medium for streamers and viewers to bi-directionally engage, co-creating a unique livestreaming experience (as P7V11 states, \emph{``the tool gave an experience that the game nor stream was able to provide''}).

However, from a self-determination perspective \cite{ryan2022self}, direct spatial input also transfers autonomy from the streamer to the audience collective, reflected in the balance of agency within the streamed content. Increased audience influence may inhibit the streamer's ability to maintain leadership or demonstrate competence \cite{youngMotivationAnalysisVideo2022}. The loss of agency could also affect how streamers attribute their performance in a games' context \cite{depping2017}, tying it towards external factors rather than their competence. The removal of autonomy from the streamer can create a level of uncertainty and unpredictability. Although uncertainty can be engaging for digital experiences (being a way to keep people interested and support serendipitous experiences \cite{kumariRoleUncertaintyMomenttoMoment2019, leongDesigningExperiencesRandomness2006}), a lack of control can also result in frustration for the streamer. When we take the perspective of streamers and viewers co-creating, then co-destruction can occur when there is a lack of trust, a lack of clear expectations, misbehaviour, and so forth \cite{jarviWhenValueCocreation2018}, which can in turn ruin the relationship between the streamer and viewers. The balance between positive effects towards social relatedness and the negative impact of loss of streamer autonomy is important to consider when using a system like VIBES or most general APGs. We consider how the maintenance of a trusting co-creation environment can be supported --- facilitating the balance of autonomy in a nuanced and controlled way to ultimately be mindful of the streamer, the viewers, and their respective needs.  

\subsection{Perspectives on Audience-Participation Games and Beyond}

Both studied applications were games, therefore, we relate our findings to audience participation games in research. Agreeing with past research \cite{Seering2017}, we found that audience members were able to take on both individual as well as collectivist roles within the studies. However, we run into similar challenges as presented by prior studies on APGs \cite{glickman2018}, namely, the issues of latency and the balance of player agency. One key finding in prior research was that text-based chat was limited in viewer expressiveness for APGs \cite{lessel2017}. Our work did improve on this aspect --- we highlighted the possibility for audience members to interact visuospatially on a visual medium (the stream), and we found that this aspect of direct spatial feedback and input was well-received. Our findings suggest that this led to higher degrees of participatory gameplay as players can directly see the outcome of their action; agreeing with prior work that visual interactions enable richer and more extensive interactions \cite{yang2020snapstream, chung2021}

VIBES contrasts existing tools such as Vispoll \cite{chung2021} and StreamSketch \cite{lu2021streamsketch}, which all allow a user to interact on a stream but do not directly impact the streamed content --- in these systems, the mediating layer of the streamers themselves decides how they want to use player input. As VIBES passes the audience input directly into the games themselves, we revisit the aspect of \emph{autonomy} within the audience collective (as it relates to self-determination theory \cite{ryan2022self}). While collective audience input enhances engagement, individual agency decreases as viewer numbers grow \cite{chenFeltEveryoneWas2024}. Feedback is important, and viewers want to feel like their actions mean something. Yet, with a large viewer base and increased individual agency, the stream may lack the social feeling of having a `crowd' \cite{chenFeltEveryoneWas2024}. Thus, challenges exist regarding individual motivators when the audience is viewed as a collective. Future work can look into building games and systems that explore individual audience identity and autonomy within a collective, perhaps being inspired by popular danmaku-participation games, which give viewers individualized control \cite{chenFeltEveryoneWas2024, wangLetsPlayTogether2023a}. 

Beyond just motivators of enjoyment and sociality, games have also been well-studied to have potential prosocial outcomes \cite{liEffectsProsocialVideo2023, whitakerRemainCalmBe2012}, which can derive from the analogue between decisions in-game and behaviour outside of it \cite{itenChoosingHelpMonsters2018a, itenDoesProsocialDecision2018}. We hypothesize that audience-participation games using direct spatial input can encourage prosocial outcomes for both the audience and the streamer, extending upon the work of Apostolellis and Bowman \cite{apostolellis2016}. Compared to usual livestreaming, which can be more passive \cite{wyndow2022subculture}, our work encodes an active visual component to encourage co-playing and co-presence. This could potentially positively affect prosocial and serious games as well, as the audience becomes more involved and feels like they have some aspect of control of that their decisions have meaning or choice. Although livestreaming has often been associated with a gaming context, its practice has been applied to many different domains, such as education \cite{chen2021} or cultural exchange \cite{lu2019}. This aligns well with serious games \cite{laamartiOverviewSeriousGames2014a}, which are games that have been used for educational purposes. Although our work has studied the context of spatial input in games contexts, future work could potentially extend these findings by leaning into studying its value in learning \cite{apostolellis2016, apostolellisExploringValueAudience2010}, especially taking advantage of visual learners.

\section{Future Work}

Our findings revealed challenges brought up in our explorations by both streamers and viewers, we expand upon challenges for improvement and possible future extensions. These recommendations aim to balance the discussed positives of creative audience interaction while being mindful of streamer agency, safety, control, and enjoyment. 

\subsection{Scalability and Streamer-Viewer Agency}
Both streamers and viewers brought up scalability as a point of concern. Streamers wondered how agency might be affected by thousands of inputs all at once, and viewers were concerned about their agency of their interactions within a large collective. To preface, the vast majority of streams on Twitch do not have particularly high viewership\footnote{\url{https://sullygnome.com/channels/14/metadata} [Last accessed: March 12, 2025]}), so designing for scale entails designing for the exception (with a distinct set of needs and requirements) rather than the norm. Scalability, and its associated effects on agency, was a dimension that we considered in design and to explore in the future, but was not as relevant towards our specific exploration. However, we speculate on the challenges associated with scale and propose future explorations and designs based on our findings.

One method of controlling the effect of scale on the balance of streamer-viewer agency, which we attempted in \emph{Terraria Interaction Mod}, was by lowering the possible number of interactions of a single individual when there is a higher total number of participants. One participant commented on this --- \emph{``I liked that the more people there were, the fewer points we get so it's not just bullying the streamer but balanced''} (P7V18). In such a case, each viewer still retains the impact of their actions, just at a less frequent rate. On the other hand, another method of reducing individual viewer agency is through aggregation of inputs, for example, in Figure 4, so that no one individual interaction can have a significant impact. Although this may dilute individual impact, it can be done more frequently, pointing towards the hypothetical trade-off between the impact of viewer action and their frequency.  

These previous suggestions presume viewers as equal, yet this is not true in existing streams --- viewers can be categorized in a hierarchical system of VIPs, mods, subscribers, etc.; leaderboards may also exist to reinforce this hierarchy. One way of balancing agency following this existing hierarchy may be to shift individual agency towards higher-tiered viewers. Furthermore, to incentivize increased engagement and interaction for viewers, some suggestions were made to allow for more individual interactions if the viewer is already engaging --- serving as a positive reward feedback system \cite{wang2011game} akin to this hierarchy. Overall, several potential mechanisms to tune VIBES at scale could be explored in future deployments; our design recommendations include dynamic interaction scaling and toggleable options for streamers to control the interaction intensity (and thus the balance of agency).

\subsection{Interaction Privacy and Anti-Harassment} 
We found that many streamers were concerned with issues regarding trolling and harassment given a new medium of audience-streamer interaction. The rise of livestreaming has inevitably led to issues with toxic behaviour in the medium \cite{enrico2022sharing, zhang2019livestreaming, rines2021conceptualizing, Groen2020}. Toxicity by viewers toward a streamer using live chat can take the form of spamming or harassment \cite{rines2021conceptualizing}, and can be challenging to deal with due to overwhelming flow \cite{enrico2022sharing}. Equivalent occurrences can take place using direct spatial input, e.g., spamming or generating unwanted inputs. To address this issue, we consider design possibilities that could incorporate proactive and reactive measures as options for safety and anti-harassment \cite{seering2017shaping}. For example, VIBES could incorporate proactive detection of spam-like inputs and prevent them from affecting the application or display interactions publically with an attached username to deter anonymous trolling; reactive measures could involve banning or timing out a troublesome user after the infraction has occurred \cite{Groen2020}. 

\subsection{Practical Deployment and Present Overhead}
In our work, VIBES is a prototype that we use as a technology probe; a more formalized deployment of such a system in practice still needs several steps that were not performed in our present implementation. For example, our Twitch extensions were deployed in the testing stages, meaning that each viewer and streamer had to be individually added to the extension as a tester rather than it being publicly available to all; moving to a public deployment requires review from Twitch themselves. This finalized deployment of the Twitch extension would likely provide several usability improvements, e.g. increased responsiveness through caching, standardized distribution channels, etc. 

At present, VIBES currently has an inevitable installation overhead at the moment due to the browser extension, which needs to be installed by the viewers to collect and pass mouse events. Because of the nature of the present architecture, there is some set-up for both the streamer and the viewer. In the future, we recommend working toward collaborating with streaming platforms for native mouse-event tracking, allowing for more streamlined public access and integration. This may also pave the way for cross-platform accessibility --- we envision that direct spatial input can be developed for and assessed for other mediums and interaction events, e.g. swipe and touch on mobile devices. 

\section{Limitations}

We reiterate that our research is largely exploratory, focusing on probing into the research questions regarding viewer and streamer experiences. Still, we identify and discuss several limitations in our study. Despite the open-ended nature of our research questions, our technology probe focused on a specific platform (Twitch) on a specific medium (browser-based computers). Future work could investigate mobile integration as well, considering the popularity of mobile viewership \cite{cunningham2019china, chrkachol2018live} and the differences in viewership behaviour \cite{dasilva2021mobile}. Although our mouse-based interactions would map naturally to touch inputs in mobile implementation from a technical perspective, we emphasize the need to extend the study of engagement and participation, especially given differences in screen size, haptic feeling, etc \cite{adepuComparisonPerformancePreference2016, wuTouchClickEffect2024}. Furthermore, while the research question was application-agnostic, our application investigation focused primarily on two specific gaming contexts. Although many livestreams do involve gameplay elements, future research could consider direct spatial input in other contexts, such as education \cite{chen2021} or cultural exchange \cite{lu2019}.  
We highlight limitations regarding participant demographics. Firstly, the streamer participants in the initial formative exploration were all low-viewer, non-full-time streamers. Although the viewership numbers are expected --- the vast majority of streamers on Twitch do have a relatively low number of viewers (stream viewership with the same order of magnitude as our deployment study is already quite rare\footnote{\url{https://sullygnome.com/channels/14/metadata} [Last accessed: March 12, 2025]}). Still, it would be important to see how direct spatial input might affect streamers who stream on a more regular basis for prolonged periods to higher viewership, especially as we highlighted scalability as a key issue to address and explore in the future. Streamer experience would be an additional factor to consider in understanding how streamers build and engage their community. In addition, although the number of participants (especially streamers) was deemed adequate for initial exploration, future work could expand our work with a more diverse and larger participant sample for increased generalizability and information power \cite{malterudSampleSizeQualitative2016b}. 

In the livestream study sessions, we recruited viewer participants who had varied levels of experience with livestream viewership; participants likely individually brought some preconceived experience towards livestreaming when answering the research questions, which ask to compare the stream with VIBES to a hypothetical stream without it. We acknowledge that in the study, we primarily relied on a comparison of a stream with VIBES to a generic recollection of the experience of watching streams, which can differ with experience. Furthermore, we also highlight limitations regarding our choice of recruiting assigned audiences rather than using existing audience members. Although this recruiting approach has been applied in prior relevant research \cite{chung2021, hamilston2016, chenIntegratingMultimediaTools2019} and offers practical methodological benefits such as controlling the level of viewership and managing scheduling, we acknowledge it may affect the nature of interactions and relationship-building between the viewers and streamers; future works could investigate deploying towards a streamers' existing audience, who also may offer a better baseline for comparison. Lastly, to extend the qualitative data captured in this work, future work could look into capturing and comparing the more quantitative metrics in psychophysiological work, such as heart rate or electrodermal activity \cite{ravajaContributionsPsychophysiologyMedia2004} to understand affective state; offering an alternative but important way to assess quality of experience \cite{vlahovicSurveyChallengesMethods2022}.  

\section{Conclusion}

In this work, we investigated the potential effects of providing audience members with real-time direct spatial input to livestreamed content, using interactive events such as click and motion on the streamed video. To facilitate this investigation, we conducted a technology probe using VIBES, our developed system that allows viewers to pass mouse events as input to affect livestreamed content on Twitch. We demonstrated the system to streamers in an initial formative exploration to understand the potential of the concept, where and how it may be useful, and what challenges might exist. We then developed two complete applications that we deployed in livestreams with viewers. We found that VIBES was well-received --- viewers and streamers alike were pleased with its ability to extend present interactions, promote streamer-audience engagement, and form collaborative participatory spaces. We reflected on VIBES within the sphere of livestreaming research and outlined present limitations and potential future work for improvement.

%
\begin{acks}
This work was supported in part by the Natural Science and Engineering Research Council of Canada (NSERC) under Discovery Grant RGPIN-2019-05624.
\end{acks}


\renewcommand{\bibnumfmt}[1]{[#1]}%
\bibliographystyle{ACM-Reference-Format}
\bibliography{sample-base}


\end{document}